\documentclass[structabstract]{aa}  
\usepackage{balance}
\usepackage{wrapfig}
\usepackage{latexsym}                                   
\usepackage{longtable}
\usepackage{graphicx}
\usepackage{natbib}
\usepackage{natbib}
\bibpunct{(}{)}{;}{a}{}{,}
\usepackage{txfonts}
\defcitealias{stonkute12}{Paper~I}
\defcitealias{stonkute13}{Paper~II}
\defcitealias{zenoviene14}{Paper~III}
\begin{document}

   \title{Stellar substructures in the solar neighbourhood}

\subtitle{IV. Kinematic Group 1 in the Geneva-Copenhagen survey}

   \author{R. \v{Z}enovien\.{e}\inst{1},
          G. Tautvai\v{s}ien\.{e}\inst{1},
          B. Nordstr\"{o}m\inst{2,3},
          E. Stonkut\.{e}\inst{1},
          \and
          G. Barisevi\v{c}ius\inst{1}
          }
   \institute{Institute of Theoretical Physics and Astronomy, Vilnius University,
              A. Gostauto 12, 01108 Vilnius, Lithuania\\
              \email{[renata.zenoviene;grazina.tautvaisiene;edita.stonkute;gintaras.barisevicius]@tfai.vu.lt}
         \and
       Dark Cosmology Centre, Niels Bohr Institute, Copenhagen University, Juliane Maries Vej 30, DK-2100, Copenhagen, Denmark
       \and
       Stellar Astrophysics Centre, Department of Physics and Astronomy, Aarhus University, Ny Munkegade 120, DK-8000 Aarhus C, Denmark \\
             \email{birgitta@nbi.ku.dk}
             }

  \date{Received September XX, 2014; accepted October XX, 2014}

 
  \abstract
   {A combined study of kinematics and chemical composition of stars is one 
of the most promising tools of research in Galaxy formation. The main goal in this field of research is to reconstruct the formation history 
of our Galaxy, to reveal the origin of the thick disc, and to find remnants of ancient mergers.}
   {We determine detailed elemental abundances in stars belonging to the so-called Group~1 of the 
Geneva-Copenhagen survey (GCS) and 
   compare the chemical composition with the Galactic thin- and thick-disc stars, with the GCS Group~2 and 
Group~3 stars, as well as with several kinematic streams of similar metallicities. The aim is to search for 
   chemical signatures that might give information 
   about the formation history of this kinematic group of stars.}
   {High-resolution spectra were obtained with the Fibre-fed Echelle 
Spectrograph (FIES) spectrograph at the Nordic Optical Telescope, La Palma, 
and were analysed
with a differential model atmosphere method. Comparison stars were observed and analysed with the same method.}
   {The average value of [Fe/H] for the 37 stars of Group~1 is $-0.20\pm 0.14$~dex. 
Investigated Group~1 stars can be separated into three age subgroups. Along with the 
main 8- and 12-Gyr-old populations,  a subgroup of stars younger than 5~Gyr can be separated as well. 
 Abundances of oxygen, $\alpha$-elements, and r-process dominated 
elements are higher than in Galactic thin-disc dwarfs. This elemental abundance pattern 
   has similar characteristics to that of the Galactic thick disc and differs slightly from those in Hercules, 
Arcturus, and AF06 stellar streams.}
  {The similar chemical composition of stars in Group~1, as well as in Group~2 and 3, with that in stars of 
the thick disc might suggest that their 
formation histories are linked. The chemical composition pattern together with the kinematic properties and ages of stars 
in the investigated GCS groups provide evidence of their common origin and possible relation 
to an ancient merging event.  A gas-rich satellite merger scenario is proposed as the most likely origin.
}


\titlerunning{Stellar substructures in the solar neighbourhood.IV.}
\authorrunning{R. \v{Z}enovien\.{e} et al. }

\maketitle


\section{Introduction}

In this series of papers (\citealt[hereafter Paper~I and Paper~II] {stonkute12, stonkute13}; and 
\citealt[hereafter Paper~III] {zenoviene14}), we investigate the detailed chemical
 composition of stellar kinematic groups that 
were identified in the Geneva-Copenhagen survey (GCS, \citealt{nordstrom04}) and were suggested to belong to 
remnants of ancient merger events in our Galaxy 
(\citealt{helmi06}). 

The formation and evolution of the Milky Way galaxy is one of the greatest outstanding questions of astrophysics.
 According Lambda cold dark matter
($\Lambda$CDM) cosmological model, large galaxies like our Galaxy emerge as an end-point of hierarchical 
 clustering,
merging, and accretion, but  we still lack the detailed physical picture of how individual stellar populations can be 
 associated with elements of the proto-cloud, 
and how different Galactic components formed and evolved. 
Even the thick disc, as a unique Galactic component discovered more than 30 years ago \citep{gilmore83}, still 
has no approved formation scenario. 
Separation of the Galactic disc into "thick" and "thin" disc populations refers to different types of stars. Thick-disc stars 
 are essentially older and more highly enriched 
in $\alpha$-elements than thin-disc stars (e.g. \citealt{soubiran05}). The enhancement in $\alpha$-elements suggests 
that the thick-disc stars were formed on relatively 
short timescales ($\sim1$~ Gyr), thus offering us a hint to the formation history of our Galaxy \citep{kordopatis13a}. 
Ancient minor mergers of satellite galaxies are considered as one possible scenario 
of the thick-disc formation. 

Numerical simulations of merger events have shown that such debris streams survive as coherent structures over 
 gigayears (\citealt{helmi04, law05, penarrubia05}). 
Stellar streams may be discovered as overdensities in the phase space distribution of stars in the solar vicinity. 
 Examining the GCS catalogue, \citet{helmi06} 
looked for stellar streams in a space of orbital apocentre, pericentre, and \textit{z}-angular momentum 
($L_{\rm z}$), the so-called APL-space. In this kind of  space 
stellar streams cluster around lines of constant eccentricity. They found three new coherent stellar groups  (Group~1, 2, and 3) with 
 distinctive ages, metallicities, and kinematics, and 
suggested that those groups might correspond to remains of disrupted satellites.

\citet{arifyanto06}, \citet{dettbarn07}, and \citet{klement08} proposed an alternative projection of the 
 eccentricity  and $L_{\rm z}$ space, which according to  
\citet{dekker76} theory of Galactic orbits can be approximated by $(U^{2} + 2V^{2})^{1/2}$ and \textit{V} 
 velocities, respectively ($U$ is pointing towards the Galactic center, $V$ into the direction of Galactic rotation). The assumption is that stars in the same 
stellar stream move on orbits that stay close together, which is justified by numerical simulations of satellite 
 disruptions \citep{helmi06}. With this method, 
\citet{arifyanto06} identified several known streams (Hyades-Pleiades, Hercules, and Arcturus) and one new, the so-called AF06 stream (later confirmed by \citealt{klement09} 
 and by \citealt{klement11}).

   \begin{figure*}
   \centering
   \includegraphics[width=\textwidth]{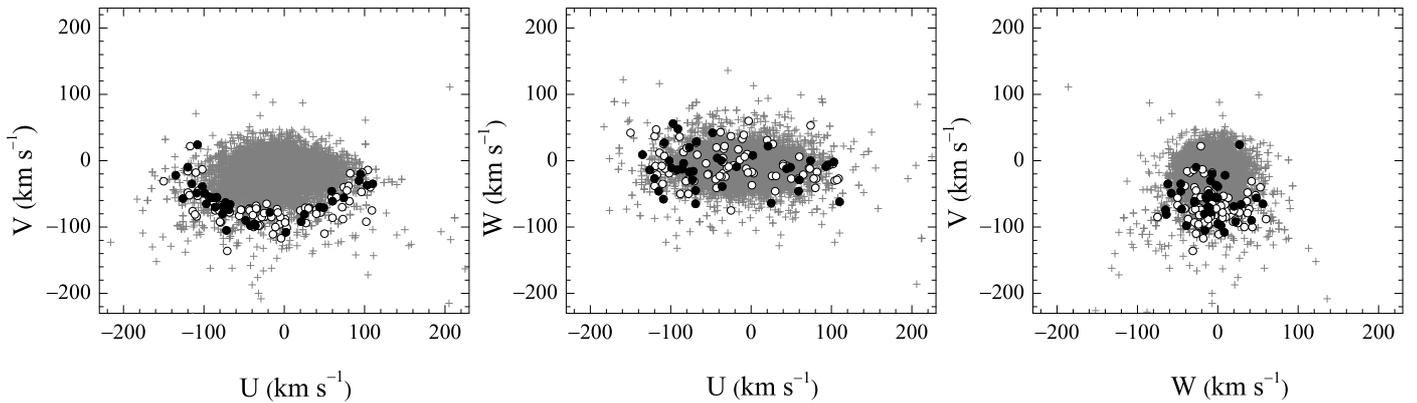}
      \caption{Velocity distribution for all stars in the sample of \citet{holmberg09} (plus signs), stars of 
Group~1 (circles), and the  stars investigated here (filled circles). 
              }
         \label{Fig.1}
   \end{figure*}

   \begin{figure*}
   \centering
 \includegraphics[width=0.95\textwidth]{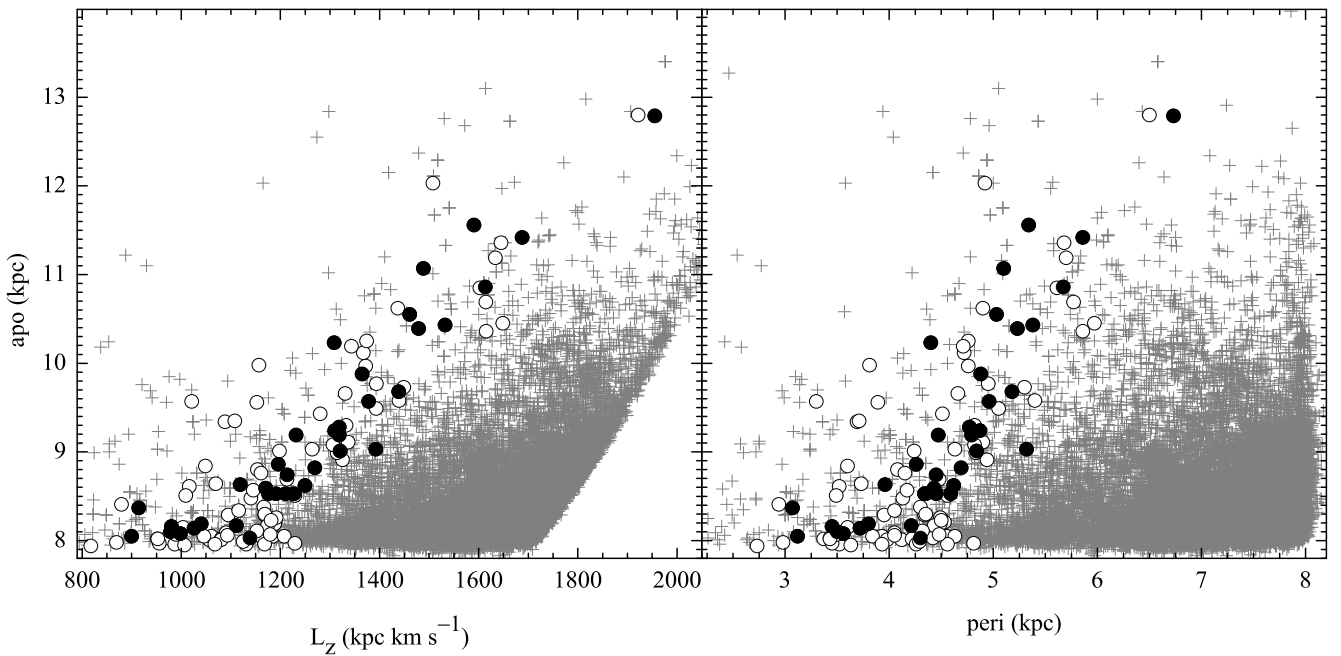}

      \caption{Distribution for the stars in the APL space. Plus signs denote the GCS sample \citep{holmberg09}, 
circles denote Group~1, the filled 
      circles are the stars we investigated. Note that these stars as well as all Group 1 stars are distributed
      in APL space with constant eccentricity. 
              }
         \label{Fig.2}
   \end{figure*}

\citet{dettbarn07} analysed the phase space distribution in a sample of non-kinematically selected low metallicity stars in the solar vicinity and determined the orbital parameters of several halo streams. 
One of those streams seemed to have precisely the same kinematics as the Sagittarius stream.

\citet{klement08} searched for stellar streams or moving groups in the solar neighbourhood, using data 
 provided by the first RAVE public data release. 
They estimated overdensities related to the Sirius, Hercules, Arcturus, and  Hyades-Pleiades moving groups. Besides,
 they found a new stream candidate 
(KFR08), suggesting that its origin is external to the Milky Way's disc. The KFR08 stream later was confirmed by 
 \citet{klement09} and \citet{bobylev10}. 
\citet{antoja12} also used the RAVE data and discovered a new group at $(U, V ) = (92, -22)\,{\rm km\,s}^{-1}$ 
in the solar neighbourhood. The new group was detected as a significant 
overdensity in the velocity distributions using a technique based on the wavelet transform. 

A mechanism, known as 'ringing' was introduced by \citet{minchev09}. \citet{minchev09} showed that the sudden energy kick imparted by the gravitational potential 
of the satellite as it crosses the plane of the disc 
can strongly perturb the velocity field of disc stars located in local volumes such as the solar neighbourhood. 
These perturbations can be observed in the 
($U,V$)  plane as arc-like features travelling in the direction of positive $V$ (see \citealt{gomez12}). 
Using this technique \citet{minchev09} confirmed presence of several known moving 
groups and predicted four new features at $V \sim -140, -120$, 40, 60~km\,s$^{-1}$. By matching the number and positions of the observed streams, they estimated that the Milky Way disc was strongly perturbed about 1.9~Gyr ago. 

Various studies have been searching for evidence of minor-merger events in the Milky Way (see review by 
 \citealt{klement10}). Large and deep chemodynamical 
surveys such as $Gaia$ \citep{perryman01} will enable us to characterise the Galaxy more precisely and to 
 examine its formation scenarios as well as the kinematic stellar group identification methods. 

An indispensable method to check the origin of various kinematic streams is the investigation of their chemical composition.  In this paper, we present the detailed chemical composition of stars belonging to Group~1 
of the Geneva-Copenhagen survey, which was identified by \citet{helmi06}. 

High-resolution spectra of 21 stars in the most metal-deficient GCS Group~3 were investigated in Paper~I 
and Paper~II. Paper~III presented results of the analysis of 32 stars in the GCS Group~2. Galactic thin-disc stars were analysed for a comparison as well. 
We present the detailed chemical composition results  for Group 1, the last group of GCS. 
This kinematic group, which  is the most numerous and metal-rich among the GCS groups, consists of 120 stars. According to the \citep{nordstrom04} catalogue, its 
mean photometric metallicity, [Fe/H], is about  
$-0.45$~dex. The Group~1 stars are distributed into two age populations of 8~Gyr and 12~Gyr. Group~1 also 
differs from the other two groups by different 
metallicity and kinematics, and tends to have a significantly smaller $z$-velocity dispersion. 
We investigated chemical composition for 37  Group~1 stars, eight comparison Galactic thin-disc stars, and five 
comparison Galactic thick-disc stars. We determined 
abundances of up to 22 chemical elements and compared them with the Galactic 
disc abundance pattern.

In Fig.~\ref{Fig.1}, we show the velocity distribution of the Galactic disc stars from \citet{holmberg09}. 
Stars belonging to GCS Group~1 
are marked with open and filled circles (the latter are used to mark stars we investigated). An average error of stellar space motions in each component  ($U$, $V$, and $W$) is 1.5~ km\,s$^{-1}$  (\citealt{nordstrom04}).
It is clear from Fig.~\ref{Fig.1} that the distribution of Group~1 
stars in the velocity space differs in comparison to other stars of the Galactic disc. For example in the 
($U,V$) plane Group~1 stars are distributed in 
a banana shape, whereas the thin-disc stars define a centrally concentrated clump. These characteristic patterns 
are the same as those of infalling dwarf satellites after several gigayears (see \citealt{helmi06} for a comprehensive discussion). In Fig.~\ref{Fig.2}, the stars are shown in the APL-space. Kinematic group stars 
spread out in the edge of the distribution of data points. The location of a star in the APL space is very accurate. A limited knowledge of the form of the Galactic potential, used to determine values of apocentre and pericentre, does not affect the distribution of points in the APL space much because the volume probed by the GCS sample is so small
that the Galactic potential is close to a constant inside this region (c.f. \citealt{helmi06}). This implies 
that the energy, and hence the orbital parameters or
the location of a star in the APL space, are determined mostly by its kinematics rather than by its spatial location (or the Galactic
potential). Changes in the Galactic potential produce only small variations in the orbital parameters. According to \citet{helmi06}, in the case of the
potential proposed by \citet{flynn96}, which was used for the orbit integrations in \citet{nordstrom04} and the following papers, a typical change in the apocentre and pericentre is of about 1--2\,\%.   
  
Despite the fact that the kinematic group members, 
over time, were dispersed through the Galactic disc, their chemical composition should remain unchanged. 
The high-resolution spectroscopic analysis is an important 
supplemental method in revealing and confirming the origin and history of the GCS stellar groups.

\begin{table*}                                                                                                                                                                                                    
        \centering                                                                                                                                                                                                
\begin{minipage}{150mm}                                                                                                                                                                                           
\caption{Parameters of programme and comparison stars}                                                                                                                                                        
\label{table:1}                                                                                                                                                                                                   
\begin{tabular}{lcrrrrrrrcrrr}                                                                                                                                                                                    
\hline                                                                                                                                                                                                            
\hline                                                                                                                                                                                                            
         Star &  Sp. type & Age & Age&   $M_{\rm V}$ & $d$ & $U$ &  $V$ &  $W$ & $e$ & $z_{\rm max}$ &  $R_{\rm peri}$ & $R_{\rm apo}$  \\                                                                
                        &        & H09$^*$  & C11$^*$ & mag           & pc & km s$^{-1}$   &  km s$^{-1}$ & km s$^{-1}$  &        & kpc & kpc & kpc \\                                                                               
\hline                                                                                                                                                                                                            
\multicolumn{13}{c}{Group 1 stars}   \\                                                                                                                                                                           
\hline                                                                                                                                                                                                            
\object HD 3795         &       K0V     &10.6   &   9.9 & 3.84  &       29      &       -48     &       -90     &       42      &       0.37    &       1.07    &       3.80    &       8.19    \\                
\object HD 4607         &       F5      &4.2    &   3.5 & 3.08  &       85      &       -108    &       24      &       27      &       0.31    &       0.86    &       6.73    &       12.79   \\                
\object HD 15777        &       G0      &11     &   11.1& 4.28  &       50      &       -77     &       -69     &       20      &       0.32    &       0.50    &       4.45    &       8.74    \\                
\object HD 22872        &       F9V     &6.2    &   5.9 & 3.59  &       74      &       -99     &       -49     &       -11     &       0.32    &       0.06    &       4.96    &       9.57    \\                
\object HD 25123        &       G0      &11     &   ...     & 4.01      &       67      &       93      &       -30     &       -9      &       0.32    &       0.03    &       5.38    &       10.43   \\        
\object HD 40040        &       G0      &12.4   &   11 & 4.12   &       65      &       44      &       -70     &       -12     &       0.31    &       0.07    &       4.45    &       8.53    \\                
\object HD 49409        &       G0V     &11.2   &   8.2 & 4.30  &       53      &       110     &       -35     &       -62     &       0.37    &       1.53    &       5.10    &       11.07   \\                
\object HD 52711        &       G4V     &7.9    &   5.1 & 4.53  &       19      &       -18     &       -78     &       -9      &       0.30    &       0.03    &       4.30    &       8.03    \\                
\object HD 60779        &       G0      &6.8    &   5.3 & 4.38  &       36      &       -126    &       -57     &       -14     &       0.40    &       0.11    &       4.40    &       10.23   \\                
\object HD 67088        &       G5      &13.7   &   ...     & 4.63      &       67      &       95      &       -20     &       -7      &       0.31    &       0.05    &       5.67    &       10.86   \\        
\object HD 67587        &       F8      &4.7    &   3.9 & 3.29  &       47      &       74      &       -56     &       0       &       0.32    &       0.12    &       4.77    &       9.28    \\                
\object HD 76095        &       G5V     &5.5    &   5.3 & 3.28  &       49      &       2       &       -108    &       8       &       0.44    &       0.25    &       3.12    &       8.05    \\                
\object HD 77408        &       F6IV    &3.7    &   4.3 & 3.52  &       50      &       -120    &       -10     &       -27     &       0.32    &       0.40    &       5.86    &       11.42   \\                
\object HD 78558        &       G0      &11.6   &   10.9& 4.44  &       37      &       -69     &       -74     &       -65     &       0.32    &       1.24    &       4.43    &       8.59    \\                
\object HD 88371        &       G2V     &13.2   &   9.4  & 4.55 &       59      &       -135    &       -22     &       9       &       0.37    &       0.31    &       5.34    &       11.56   \\                
\object HD 88446        &       F8      &7.3    &   7.8 & 3.77  &       67      &       -42     &       -98     &       4       &       0.41    &       0.17    &       3.45    &       8.16    \\                
\object HD 90508        &       G1V     &10.4   &   7.4  & 4.63 &       23      &       21      &       -92     &       22      &       0.37    &       0.51    &       3.72    &       8.14    \\                
\object HD 109498       &       G3V     &10.4   &   ...     & 4.53      &       69      &       59      &       -46     &       -46     &       0.26    &       0.79    &       5.32    &       9.03    \\        
\object HD 111367       &       G1V     &9.5    &   8.5 & 4.11  &       86      &       -69     &       -73     &       -45     &       0.33    &       0.76    &       4.34    &       8.53    \\                
\object HD 135694       &       K0      &12.8   &   6.7 & 4.82  &       71      &       -36     &       -98     &       -39     &       0.40    &       0.57    &       3.50    &       8.10    \\                
\object HD 138750       &       F8      &3.3    &   3.2 & 2.60  &       116     &       -72     &       -105    &       -16     &       0.46    &       0.18    &       3.07    &       8.37    \\                
\object HD 140209       &       G0      &10.4   &   9.6 & 4.03  &       71      &       -73     &       -63     &       -29     &       0.30    &       0.37    &       4.62    &       8.62    \\                
\object HD 149105       &       G0V     &6      &   5.8 & 3.37  &       53      &       49      &       -71     &       -10     &       0.32    &       0.06    &       4.35    &       8.53    \\                
\object HD 149890       &       F8V     &6.8    &   6.1 & 4.15  &       39      &       60      &       -61     &       -29     &       0.31    &       0.38    &       4.69    &       8.82    \\                
\object HD 156617       &       G5      &9.3    &   9.1 & 3.92  &       66      &       -102    &       -39     &       0       &       0.30    &       0.11    &       5.18    &       9.68    \\                
\object HD 156893       &       G5      &8.9    &   8.5 & 3.74  &       77      &       -97     &       -65     &       56      &       0.35    &       1.60    &       4.47    &       9.19    \\                
\object HD 157214       &       G0V     &13.9   &   8.1 & 4.60  &       14      &       25      &       -81     &       -64     &       0.32    &       1.33    &       4.21    &       8.17    \\                
\object BD +40 3374     &       K1      &...            &   6.4 & 6.43  &       49      &       -109    &       -49     &       -58     &       0.34    &       1.15    &       4.88    &       9.88    \\                
\object HD 171009       &       G5      &11.7   &   6.7 & 4.24  &       66      &       -68     &       -66     &       29      &       0.30    &       0.68    &       4.59    &       8.53    \\                
\object HD 171242       &       G0      &9.2    &   8.5 & 4.02  &       62      &       103     &       -37     &       -2      &       0.35    &       0.10    &       5.03    &       10.55   \\                
\object HD 178478       &       G5      &15.8   &   ...     & 4.90      &       47      &       -86     &       -70     &       -12     &       0.35    &       0.08    &       4.26    &       8.86    \\        
\object HD 188326       &       G8IV    & ...           &   8.7 & 3.83  &       56      &       -91     &       -56     &       48      &       0.31    &       1.24    &       4.87    &       9.24    \\                
\object HD 206373       &       G0V     &5.5    &   5.6  & 3.19 &       104     &       -37     &       -95     &       0       &       0.39    &       0.11    &       3.56    &       8.08    \\                
\object HD 210483       &       G1V     &9.3    &   6.6 & 4.03  &       51      &       -77     &       -80     &       -17     &       0.37    &       0.15    &       3.96    &       8.63    \\                
\object HD 211476       &       G2V     &7.1    &   6.7 & 4.59  &       31      &       -115    &       -35     &       -46     &       0.33    &       0.87    &       5.23    &       10.39   \\                
\object HD 217511       &       F5      &2      &   2.1 & 2.23  &       122     &       -84     &       -55     &       -4      &       0.30    &       0.08    &       4.84    &       9.01    \\                
\object HD 219175       &       F9V     &3.5    &   2.9 & 4.61  &       39      &       -92     &       -55     &       -14     &       0.31    &       0.10    &       4.79    &       9.19    \\                
\hline                                                                                                                                                                                                            
\multicolumn{13}{c}{  Thin-disc stars }  \\                                                                                                                                                                       
\hline                                                                                                                                                                                                            
\object HD 115383 &     G0V &   4.3 & 5.1 & 3.97  &     18 &    -38   & 2  &      -18             &       0.09            &       0.18            &       7.54            &        8.99    \\                        
\object HD 127334 &     G5V &   10.5& 10.7& 4.51  &     24 &    30    & -4 &       -2              &       0.12            &       0.08            &       7.14            &        9.10    \\                        
\object HD 136064 &     F9IV &  3   & 3.7 & 3.12  &     25 &    62    & -29 &       -24             &       0.23            &       0.29            &       5.83            &        9.37    \\                        
\object HD 163989 &     F6IV &  2.2 & 2.5 & 2.51  &     32 &    -29   & -27 &       -22             &       0.10            &       0.22            &       6.64            &        8.16    \\                        
\object HD 187013 &     F7V      &2.8 & ...       &     3.34  & 22 &    38    &    -8  &   -25             &       0.14            &       0.28            &        6.88            &       9.20    \\                
\object HD 187691 &     F8V  &  3.3 & 3.1 & 3.71  &     19 &    -3    & -3  &      -25             &       0.03            &       0.27            &       7.88            &        8.28    \\                        
\object HD 200790 &     F8V  &  2.2 & 2.6 & 2.51  &     49 &    19    & -37 &       10              &       0.15            &       0.27            &       6.03            &        8.21    \\                        
\object HD 220117 &     F5V  &  1.8 & 1.6  & 2.66  &    42 &    -15   & -21 &       -16             &       0.07            &       0.12            &       7.03            &        8.02    \\                        
\hline                                                                                                                                                                                                            
\multicolumn{13}{c}{      Thick-disc stars }  \\                                                                                                                                                                  
\hline                                                                                                                                                                                                            
\object HD 150433 &     G0      &13.8 &7.1      & 4.86   &      30    & -8    &    -58        &    -46         &   0.21            &       0.73            &        5.21            &       7.97    \\                
\object HD 181047 &     G8V     &13.9 &10.2     & 4.96   &      47        &      -99       &     -43        &    -17         &   0.30            &       0.15            &        5.12            &       9.57    \\
\object HD 186411 &     G0      &6    & ...     & 3.36   &      88        &      -64       &     -56        &    -6          &   0.26            &       0.02            &        4.96            &       8.52    \\
\object HD 195019 &     G3IV-V  &9.7  &8.6      & 3.96  &       39        &      -73       &     -76        &    -39         &   0.34            &       0.61            &        4.18            &       8.57    \\
\object HD 198300 &     G0      &10.3 &7.1      & 4.89  &       53        &      81        &     -20        &    -24         &   0.28            &       0.28            &        5.84            &       10.26   \\
\hline
\end{tabular}
\end{minipage}
\tablefoot{$^*$ H09 - ages taken from \citet{holmberg09}, C11 - from \citet{casagrande11}.}
\end{table*}

\section{Observations and analysis}

We obtained high-resolution spectra of the programme and comparison stars  with the Fibre-fed Echelle 
Spectrograph on the Nordic Optical 2.5~m 
telescope (NOT) during 2008, 2011, and 2012. This spectrograph gives spectra of resolving power 
($R\approx68\,000$) in the wavelength range of 
3680--7270~{\AA}. We exposed the spectra  to reach a signal-to-noise ratio higher than 100. We elected the programme 
stars  so that they were observable 
from the NOT and were brighter than $V = 10$~mag. 
Reductions of CCD images were made with the FIES pipeline $FIEStool$, which performs  a complete reduction: 
calculation of reference frame, bias and 
scattering subtraction, flat-field division, wavelength calibration and other procedures 
 (http://www.not.iac.es/instruments/fies/fiestool). A list of the observed 
stars and some of their parameters (taken from the catalogue of \citealt{holmberg09} and SIMBAD) are 
presented in Table~\ref{table:1}.

\begin{figure}
   \centering
   \includegraphics[width=0.47\textwidth]{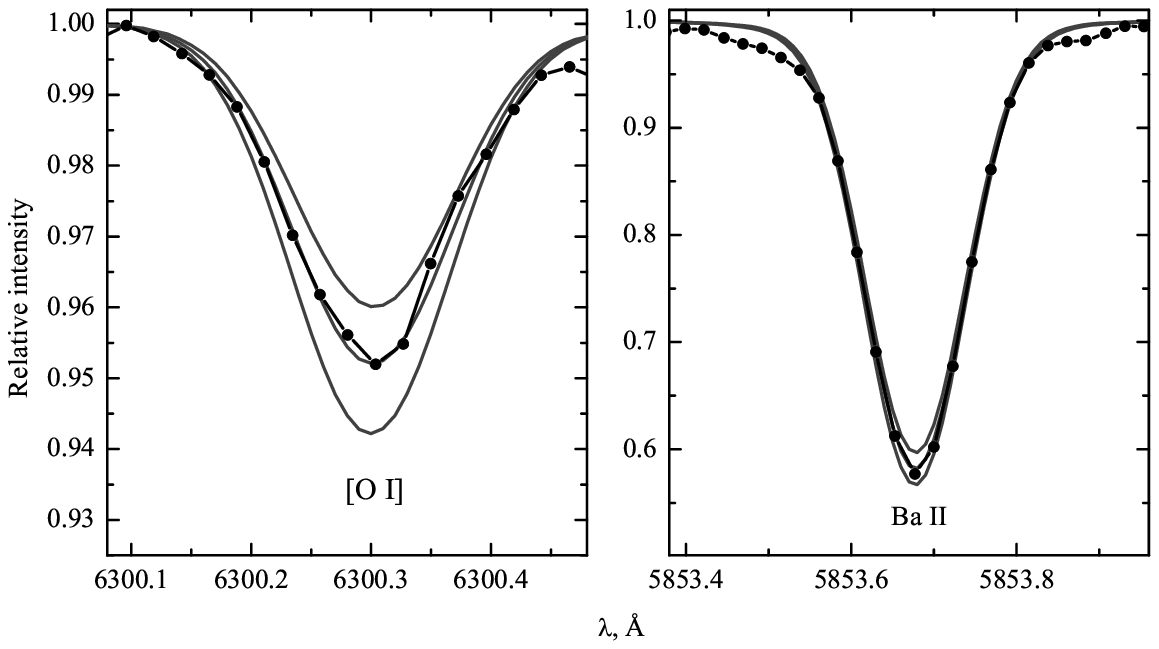}
\caption{Synthetic spectrum fit to the forbidden [O\,{\sc i}] line at 6300.3 {\AA} in the observed spectrum of 
HD\,157214 (left panel) and to the barium line at 
5853~{\AA} in the observed spectrum of HD\,52711. The observed spectra are shown by solid lines with dots. 
The dark grey solid lines are synthetic 
      spectra with ${\rm [O/Fe]}=0.42 \pm 0.1$ and with ${\rm [Ba/Fe]} = -0.08 \pm 0.1$, respectively.  } 
\label{Fig.3}
\end{figure}

\begin{figure}
   \centering
   \includegraphics[width=0.47\textwidth]{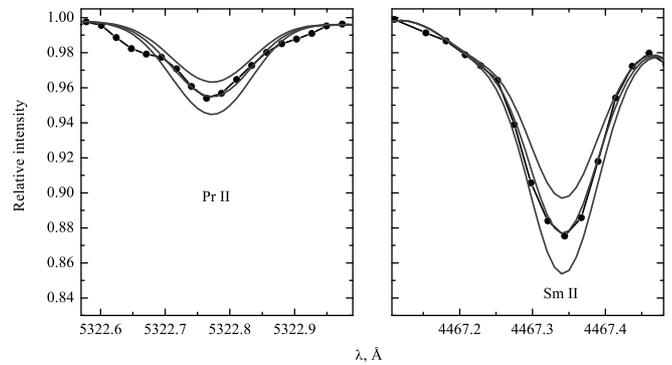}
\caption{Synthetic spectrum fit to the praseodymium line at 5323~{\AA} in the observed spectrum of HD\,3795 
(left panel) and to the samarium line at 4467~{\AA} in the observed spectrum of HD\,22872. The observed spectra 
are shown by solid lines with dots. The dark grey solid lines are synthetic spectra with 
${\rm [Pr/Fe]} = 0.54 \pm 0.1$ and ${\rm [Sm/Fe]} = 0.01 \pm 0.1$, respectively. } 
\label{Fig.4}
\end{figure}


   \begin{figure}
   \centering
   \includegraphics[width=0.47\textwidth]{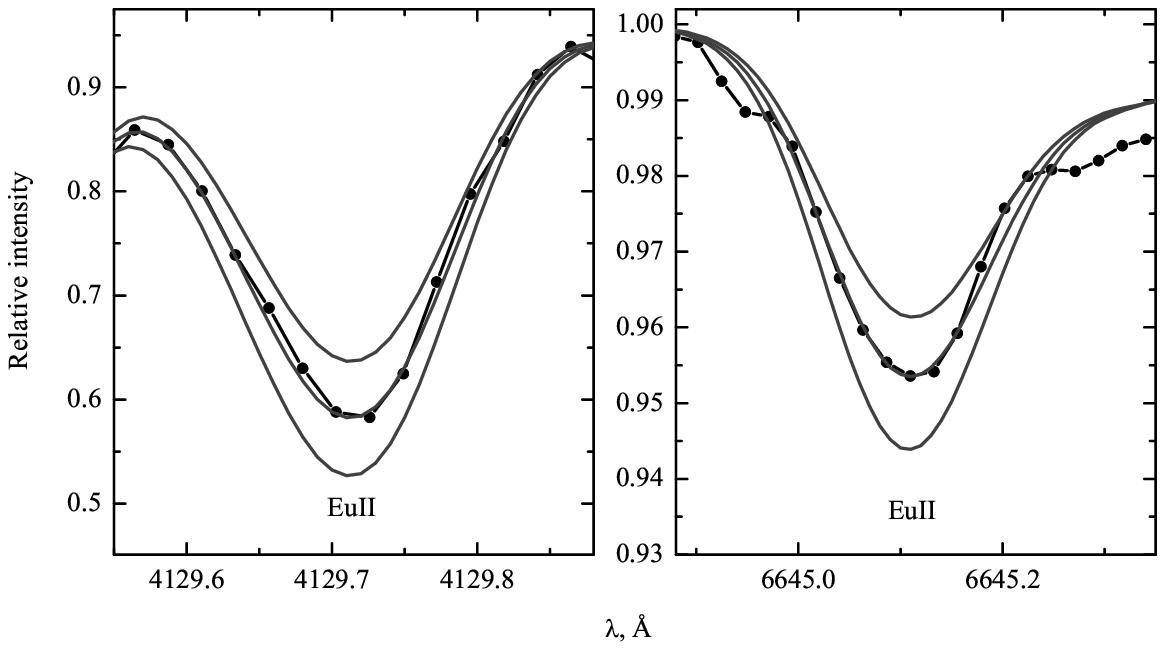}
      \caption{Synthetic spectrum fit to the europium lines at 4129~{\AA} and 6645~{\AA}. The observed spectrum for 
the programme star HD\,149105 is shown as a solid line with dots. The dark grey solid lines are synthetic spectra 
with ${\rm [Eu/Fe]} = 0.04 \pm 0.1$ and ${\rm [Eu/Fe]} = 0.11 \pm 0.1$ for these two lines, respectively. 
              }
         \label{Fig.5}
   \end{figure}

We analysed the spectra  using the differential model atmosphere technique described in Papers~I, II, and III. 
Here we will revisit only a few details. The Eqwidth and BSYN program packages, developed at the Uppsala 
Astronomical Observatory, were used to carry out the calculation of abundances from measured equivalent 
widths and synthetic spectra, respectively.
A set of plane-parallel, line-blanketed, constant-flux LTE model atmospheres \citep{gustafsson08} were 
taken from the MARCS stellar model 
atmosphere and flux library (http://marcs.astro.uu.se/).

We used the Vienna Atomic Line Data Base (VALD, \citealt{piskunov95}) to prepare input data for the calculations. We took the atomic oscillator 
strengths for the main spectral lines we analysed   from  
an inverse solar spectrum analysis performed in Kiev \citep{gurtovenko89}.

\begin{figure*}
   \centering
   \includegraphics[width=0.47\textwidth]{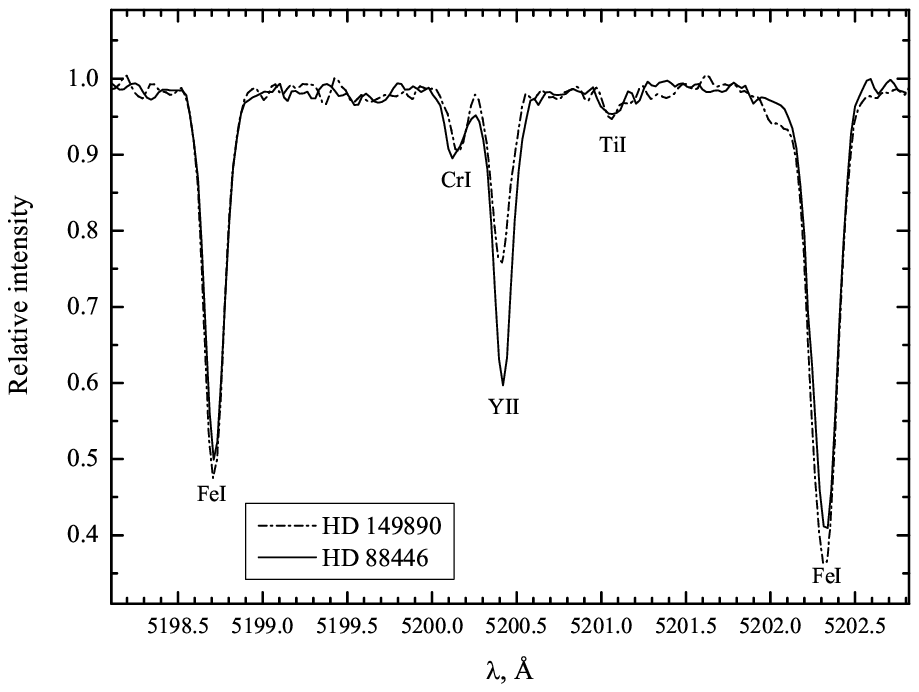}
      \includegraphics[width=0.47\textwidth]{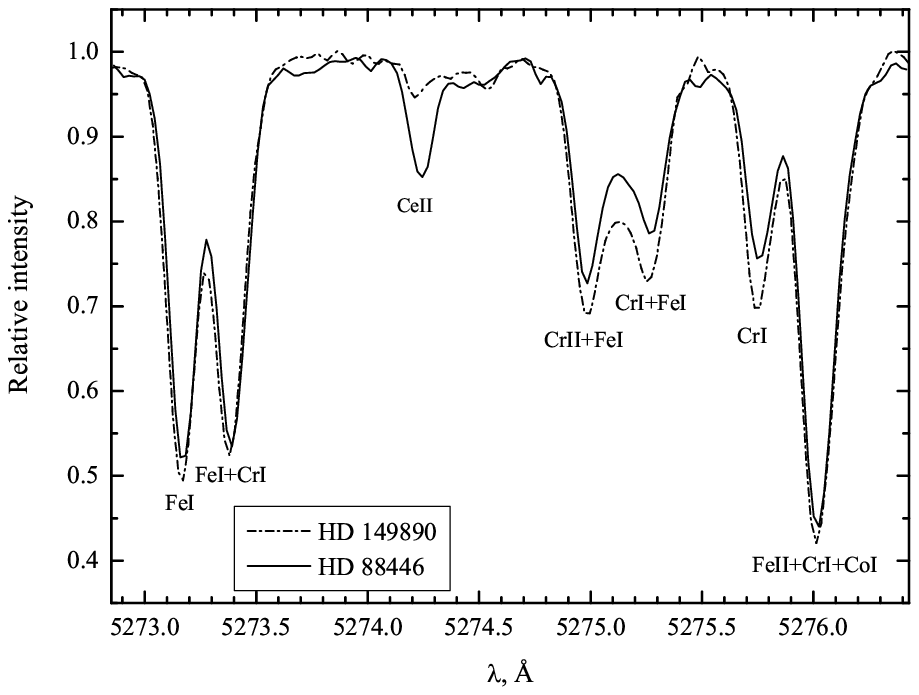}
      \caption{NOT-FIES spectra of  Group~1 stars HD\,149890 and HD\,88446. These two spectra are over-plotted 
to see the difference in spectral lines 
      of elements produced in $s$-process, while lines of other chemical elements are similar. 
              }
         \label{Fig.6}
   \end{figure*}

Initial values of the effective temperatures for the programme stars were taken from \citet{holmberg09} 
and then carefully checked and corrected, if needed, by 
forcing Fe~{\sc i} lines to yield no dependency of iron abundance on excitation 
potential through changes to the model effective temperature. 
We used the ionisation equilibrium method to find surface gravities of the programme stars 
by forcing neutral and ionised iron lines to yield the same iron abundances.
Microturbulence velocity values corresponding to the minimal line-to-line  
Fe~{\sc i} abundance scattering were chosen as correct values. 
We performed the spectral synthesis method  for the determination of oxygen, yttrium, zirconium, barium, lanthanum, cerium, 
praseodymium, neodymium, samarium, and europium abundances. 
Several fits of the synthetic line profiles to the observed spectra are shown in Figs.~\ref{Fig.5}--\ref{Fig.8}. 
We complied atomic parameters of lines in the intervals of spectral syntheses  from the VALD database.
We calibrated all log~\textit{gf} values  to fit to the solar spectrum by \cite{kurucz05} with solar abundances 
from \cite{grevesse00}.
We took hyperfine structures and isotope shifts into account as appropriate.  
Abundances of other chemical elements were determined using equivalent widths of their lines. 
We determined abundances of Na and Mg taking  non-local thermodynamical equilibrium (NLTE)  into 
account. The equivalent widths of the lines were measured by fitting of a Gaussian profile using the {\sc 4A} software 
package \citep{ilyin00}.
  
The uncertainties in abundances are due to several sources: uncertainties caused by 
analysis of individual lines, including random errors of atomic data and continuum 
placement and  uncertainties in the stellar parameters.
The sensitivity of the abundance 
estimates to changes in the atmospheric parameters by the assumed errors 
$\Delta$[El/H]\footnote{We use the customary spectroscopic notation
[X/Y]$\equiv \log_{10}(N_{\rm X}/N_{\rm Y})_{\rm star} -\log_{10}(N_{\rm X}/N_{\rm Y})_\odot$.} 
are illustrated  for the star HD\,52711  (Table~\ref{table:2}). Clearly, possible 
parameter errors do not affect the abundances seriously; the element-to-iron 
ratios, which we use in our discussion, are even less sensitive. 

The scatter of the deduced abundances from different spectral lines $\sigma$
gives an estimate of the uncertainty due to the random errors. The mean value 
of  $\sigma$ is 0.04~dex, thus the uncertainties in the derived abundances that 
are the result of random errors amount to approximately this value.


   \begin{table}
   \centering
   \begin{minipage}{80mm}
      \caption{Effects on derived abundances resulting from model changes for the star HD\,52711.} 
        \label{table:2}
      \[
         \begin{tabular}{lrrrr}
            \hline
            \hline
            \noalign{\smallskip}
Ion & ${ \Delta T_{\rm eff} }\atop{ +100 {\rm~K} }$ & 
            ${ \Delta \log g }\atop{ +0.3 }$ & 
            ${ \Delta v_{\rm t} }\atop{ +0.3~{\rm km~s}^{-1}}$ &
            ${\rm Total} $ \\ 
            \noalign{\smallskip}
            \hline
            \noalign{\smallskip}
[O\,{\sc i}]      & $0.03  $ & $0.13  $ & $0.01  $  & $0.13$ \\             
Na\,{\sc i}       & $0.05  $ & $-0.02 $ & $-0.01 $  & $0.05$ \\             
Mg\,{\sc i}       & $0.05  $ & $-0.02 $ & $-0.03 $  & $0.06$ \\             
Al\,{\sc i}       & $0.04  $ & $0.00  $ & $0.00  $  & $0.04$ \\             
Si\,{\sc i}       & $0.03  $ & $0.01  $ & $-0.01 $  & $0.03$ \\             
Ca\,{\sc i}       & $0.07  $ & $-0.01 $ & $-0.02 $  & $0.07$ \\             
Sc\,{\sc ii}      & $-0.00 $ & $0.11  $ & $-0.04 $  & $0.12$ \\             
Ti\,{\sc i}       & $0.09  $ & $0.00  $ & $-0.01 $  & $0.09$ \\             
Ti\,{\sc ii}      & $0.01  $ & $0.12  $ & $-0.05 $  & $0.13$ \\             
V\,{\sc i}        & $0.11  $ & $-0.00 $ & $-0.01 $  & $0.11$ \\             
Cr\,{\sc i}       & $0.08  $ & $-0.02 $ & $-0.06 $  & $0.10$ \\             
Cr\,{\sc ii}      & $-0.03 $ & $0.10  $ & $-0.09 $  & $0.14$ \\             
Fe\,{\sc i}       & $0.07  $ & $-0.02 $ & $-0.07 $  & $0.10$ \\             
Fe\,{\sc ii}      & $-0.02 $ & $0.11  $ & $-0.07 $  & $0.13$ \\             
Co\,{\sc i}       & $0.08  $ & $0.01  $ & $-0.01 $  & $0.08$ \\             
Ni\,{\sc i}       & $0.06  $ & $-0.00 $ & $-0.04 $  & $0.07$ \\             
Y\,{\sc ii}       & $0.02  $ & $0.10  $ & $-0.12 $  & $0.16$ \\             
Zr\,{\sc i}       & $0.11  $ & $0.01  $ & $0.01  $  & $0.11$ \\             
Zr\,{\sc ii}      & $0.02  $ & $0.13  $ & $0.01  $  & $0.13$ \\             
Ba\,{\sc ii}      & $0.06  $ & $0.08  $ & $-0.18 $  & $0.21$ \\             
La\,{\sc ii}      & $0.03  $ & $0.12  $ & $0.01  $  & $0.12$ \\             
Ce\,{\sc ii}      & $0.03  $ & $0.11  $ & $0.01  $  & $0.11$ \\             
Pr\,{\sc ii}      & $0.02  $ & $0.12  $ & $0.01  $  & $0.12$ \\             
Nd\,{\sc ii}      & $0.03  $ & $0.12  $ & $-0.01 $  & $0.12$ \\             
Sm\,{\sc ii}      & $0.04  $ & $0.11  $ & $-0.01 $  & $0.12$ \\             
Eu\,{\sc ii}      & $0.03  $ & $0.12  $ & $0.01  $  & $0.12$ \\             

            \hline
         \end{tabular}
      \]
\end{minipage}
\tablefoot{The table entries show the effects on the logarithmic abundances relative to hydrogen, $\Delta {\rm [El/H]}$. }
   \end{table}

\begin{table}
\centering
\begin{minipage}{80mm}
\caption{Comparison with previous studies.}
\label{table:3} 
\begin{tabular}{lrrrrrrrr}
\hline\hline   
Quantity         & Ours--B14 & Ours--R06 & Ours--E93 \\
\hline
$T_{\rm eff} $  & $ -31 \pm 48 $    & $74 \pm 51 $     & $-12 \pm 65 $     \\
log $g$         & $ -0.18 \pm 0.10 $ & $-0.16 \pm 0.15 $ & $ -0.28 \pm 0.24 $ \\
${\rm [Fe/H]}$  & $ 0.04 \pm 0.05 $ & $ 0.09 \pm 0.06 $ & $ 0.15 \pm 0.06 $         \\
${\rm[Na/Fe]}$  & $ -0.01 \pm 0.03 $ & $ -0.05 \pm 0.05 $ & $ -0.05 \pm 0.07 $   \\
${\rm[Mg/Fe]}$  & $ 0.05 \pm 0.08       $ & $ -0.07 \pm 0.05 $ & $ 0.06 \pm 0.11  $  \\
${\rm[Al/Fe]}$  & $ 0.00 \pm 0.06 $ & $ -0.02 \pm 0.06 $ & $ -0.11 \pm 0.05 $   \\
${\rm[Si/Fe]}$  & $ 0.01 \pm 0.03 $ & $ -0.03 \pm 0.03 $ & $ -0.02 \pm 0.03 $   \\
${\rm[Ca/Fe]}$  & $ 0.02 \pm 0.03 $      & $ 0.04 \pm 0.02 $ & $ 0.03 \pm 0.05 $  \\
${\rm[Sc/Fe]}$  & $ ...   $         & $ 0.01 \pm 0.07 $ & $ ... $  \\
${\rm[Ti/Fe]}$  & $ 0.02 \pm 0.04 $      & $ 0.06 \pm 0.04 $ & $ -0.04 \pm 0.06 $   \\
${\rm[V/Fe]}$   & $ ... $           & $ 0.03 \pm 0.05 $ & $ ...  $ \\
${\rm[Cr/Fe]}$  & $ 0.01 \pm 0.02 $      & $ 0.02 \pm 0.03 $ & $  ... $  \\
${\rm[Co/Fe]}$  & $ ... $         & $ -0.01 \pm 0.04 $ & $ ...  $  \\
${\rm[Ni/Fe]}$  & $ 0.00 \pm 0.03 $ & $ -0.02 \pm 0.02 $ & $ -0.12 \pm 0.05  $  \\
${\rm[Y/Fe]}$   & $ -0.05 \pm 0.06 $ & $ -0.10 \pm 0.09 $ & $ -0.12 \pm 0.09 $  \\
${\rm[Ba/Fe]}$  & $ -0.02 \pm 0.14 $ & $ 0.06 \pm 0.13 $ & $ -0.06 \pm 0.18  $ \\
${\rm[Ce/Fe]}$  & $ ... $           & $ 0.01 \pm 0.15 $ & $ ... $  \\
${\rm[Nd/Fe]}$  & $ ... $           & $ -0.06 \pm 0.10 $ & $ ... $    \\
${\rm[Eu/Fe]}$  & $ ...  $          & $ -0.02 \pm 0.11 $ & $ ... $   \\
\hline
\end{tabular}
\end{minipage}
\tablefoot{Mean differences and standard deviations of the main parameters and abundance ratios [El/Fe] for
 13 stars of Group~1 in common with \citet[B14] {bensby14}, 7 stars of Group~1 in common with 
\citet[R06] {reddy06}, and 8 thin-disc stars in common with \citet[E93] {edvardsson93}.}
\end{table}

Effective temperatures for all  stars investigated here are also available in
\citet{holmberg09} and \citet{casagrande11}.
\citet{casagrande11} provided astrophysical parameters for the Geneva-Copenhagen survey by applying the 
infrared flux method to determine the effective temperature.
In comparison to \citet{holmberg09}, stars in the catalogue of \citet{casagrande11} are 
on average 100~K hotter. For the stars investigated here, our spectroscopic temperatures are on average only $40\pm 70$~K 
hotter than in \citet{holmberg09} and $40\pm 90$~K cooler than in \citet{casagrande11}.  
The [Fe/H] values for all of the stars we investigated are available in \citet{holmberg09} as well as in \citet{casagrande11}. 
A comparison between \citet{holmberg09} and \citet{casagrande11} shows that the latter gives [Fe/H] values that are 
more metal-rich on average by 0.1~dex.
For our programme stars, we obtain a difference of $0.1\pm 0.1$~dex in comparison with \citet{holmberg09} and we obtain no 
systematic difference, but a scatter of 0.1~dex, in comparison with \citet{casagrande11}.  The same result was found in 
comparing the atmospheric parameters determined for Group~3 stars in our \citetalias{stonkute12} and for Group~2 
in  \citetalias{zenoviene14}.

Some stars from our sample were previously investigated by other authors. In Table~\ref{table:3}, we present a comparison 
with results by \citet{bensby14} 
and \citet{reddy06}, who investigated several stars in common with our work.
Eight thin-disc stars that we investigated for a comparison have been analysed previously by
\citet{edvardsson93}. 
 Slight differences in the log\,$g$ values lie within the errors of uncertainties and are caused mainly by 
differences in the applied determination methods. We see that titanium and zirconium abundances 
determined using both neutral and ionised 
lines agree well and confirm the log\,$g$ values determined using iron lines. Overall, our [El/Fe] for the stars in common 
agree very well with those in other studies.

\section{Results and discussion}

The atmospheric parameters, $T_{\rm eff}$, log\,$g$, $v_{t}$, [Fe/H] and the abundances of 21 chemical elements relative 
to iron [El/Fe] of the programme and comparison stars are presented in Tables~4 and 5. The number of lines and 
the line-to-line scatter ($\sigma$) are presented as well.

  \begin{figure*}
   \centering
   \includegraphics[width=0.85\textwidth]{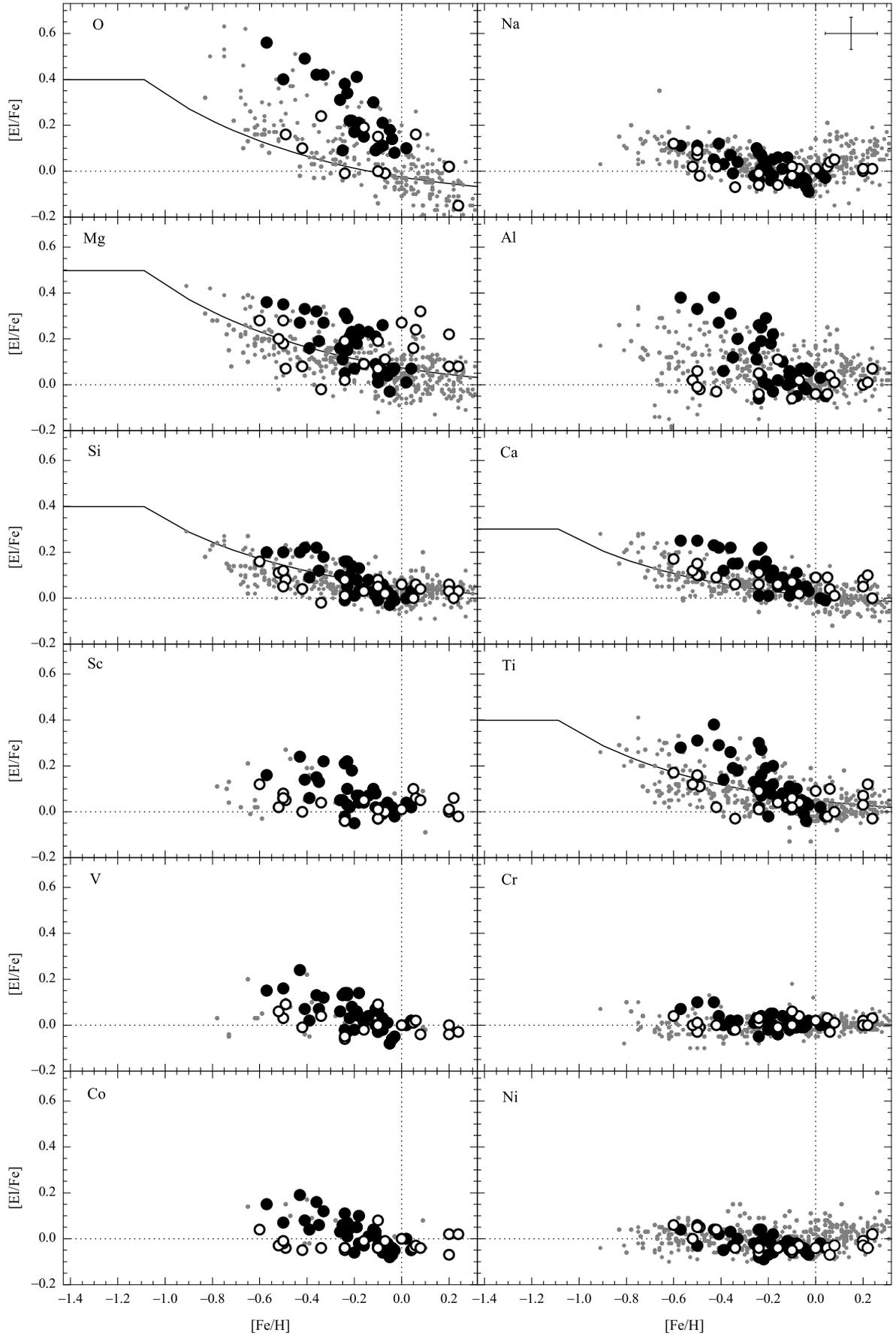}
     \caption{[El/Fe] ratio as a function of [Fe/H] for Group~1 stars (filled circles) investigated here and for  comparison 
thin-disc stars analysed in this work,  
     \citetalias{stonkute12}, and \citetalias{zenoviene14} (open circles). 
The data for the Milky Way thin-disc dwarfs 
     taken from other studies are shown as a grey dots. Solid lines are Galactic thin-disc chemical evolution models 
presented by \citeauthor{pagel95} (1995). Average uncertainties are shown in the box for Na.}
       \label{Fig.7}
   \end{figure*}
   
  \begin{figure*}
   \centering
   \includegraphics[width=0.85\textwidth]{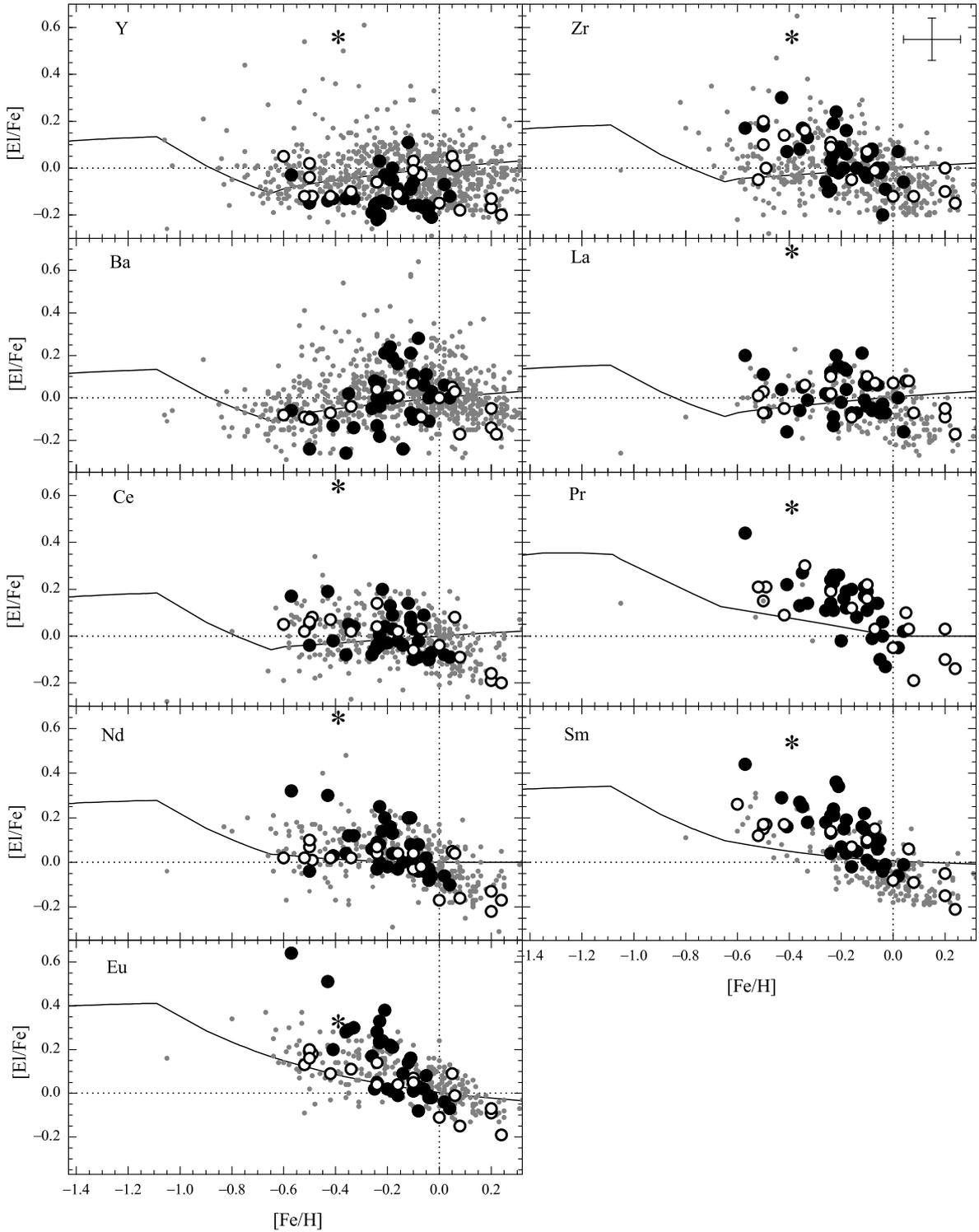}
     \caption{[El/Fe] ratio as a function of [Fe/H] for Group~1 stars (filled
circles) investigated here and for comparison 
thin-disc stars analysed in this work,  
     \citetalias{stonkute13}, and  \citetalias{zenoviene14} (open circles). The s-process enhanced star (HD~88446) 
is marked as an asterisk. 
     Grey dots correspond to the data for the Milky Way thin-disc dwarfs taken from other studies. 
The Galactic thin-disc chemical evolution model is shown 
     as a solid line (\citealt{pagel97}). Average uncertainties are shown in the box for Zr.}
       \label{Fig.8}
   \end{figure*}

The results are graphically displayed in Figs.~\ref{Fig.7} and \ref{Fig.8}.  We display elemental abundance 
ratios of Group~1 stars together with 
data of thin-disc stars  investigated here and in Papers~I, II, and III, as well as with results taken from 
other thin-disc studies (\citealt{edvardsson93, 
gratton94, koch02, bensby05, reddy06, zhang06, brewer06, mashonkina07}, and \citealt{mishenina13}). 
The chemical evolution models of the 
thin-disc were taken from \citet{pagel95, pagel97}. 
The thin-disc stars from \citet{edvardsson93} and \citet{zhang06} were selected  using the membership 
probability evaluation method described by \citet{trevisan11},
since their lists contained stars of other Galactic components as well. The same kinematic approach in assigning 
thin-disc membership was used in  
\citet{bensby05} and \citet{reddy06}, which means that the thin-disc stars used for  comparison are uniform 
in that respect. 

One star in Group~1 is rich in elements predominantly produced in s-process. As shown in Table~5 and 
Figs.~\ref{Fig.6} and~\ref{Fig.8}, HD~88446 has much stronger 
lines of elements predominantly produced in s-process and consequently much 
higher abundances of these elements. According to the definition of \citet{beers05}, HD~88446, with 
its ${\rm [Ba/Fe]}=1.04$ and ${\rm [Ba/Eu]}=0.70$,  falls in the category of the s-process-rich stars.

The metallicity of  Group~1 stars we investigated lie in  a broad interval of $0.04 \geq {\rm [Fe/H]} \geq -0.57$ with the average [Fe/H] equal to $-0.20\pm 0.14$~dex. Abundances of chemical 
elements are rather homogeneous and show similar overabundances of $\alpha$-elements and r-process-dominated 
chemical elements with respect to thin-disc stars, as we also found for the stars of GCS Group~2 and 3.  
This elemental abundance pattern has similar characteristics with that in the Galactic thick-disc.

\subsection{Comparison with the thick-disc} 

In Table~\ref{table:6} we present a comparison of mean [El/Fe] ratios 
calculated for stars of Group~1 and thick-disc stars at the same metallicity interval $-0.57 < {\rm [Fe/H]} < 0.04$. 
Twenty-six thick-disc stars in this 
metallicity interval were investigated by  \citet{bensby05}, 37 stars by \citet{reddy06}, 10 stars by 
\citet{mashonkina07}, 51 stars by \citet{stanford12}, 
and 7 stars by \citet{mishenina13}. When comparing oxygen abundances, we did not use the results reported 
by \citet{reddy06} and \citet{stanford12} 
because they investigated the O\,{\sc i}  line, while we studied [O\,{\sc i}]. The studies by \citet{mashonkina07} 
and \citet{mishenina13} were included 
in the comparison to enlarge the information on neutron capture elements.  The average values of $\alpha$-element 
abundances included Mg, Si, and Ca. 
Titanium was excluded because this element was not determined in one of the studies (\citealt{mishenina13}). 
The comparison shows that the deviations do not exceed the uncertainties. 
 
 \begin{table}
 \setcounter{table}{5}
   \centering
   \begin{minipage}{80mm}
      \caption{Comparison with thick-disc studies.} 
        \label{table:6}
      \[
         \begin{tabular}{lrrrrr}
            \hline
            \hline
            \noalign{\smallskip}
[El/Fe] & ${\rm Ours-}\atop {\rm B14}$ & 
            ${\rm Ours- }\atop{\rm R06}$ & 
            ${\rm Ours- }\atop{\rm Ma07}$ &
            ${\rm Ours- }\atop{\rm S12} $ & 
            ${\rm Ours- }\atop{\rm Mi13} $ \\
            \noalign{\smallskip}
            \hline
            \noalign{\smallskip}
${\rm[O/Fe]}$        & $-0.02   $ & $...   $ & $...   $ & $ ...$ & $ 0.08$\\
${\rm[Na/Fe]}$          & $-0.02   $ & $-0.06   $ & $...   $ & $ -0.02$ & $ ...$\\
${\rm[Mg/Fe]}$       & $-0.03  $ & $-0.06  $ & $...   $ & $ -0.01$ & $ -0.03$\\
${\rm[Al/Fe]}$          & $-0.07   $ & $-0.12   $ & $...   $ & $ -0.07$ & $ ...$\\
${\rm[Si/Fe]}$       & $-0.04   $ & $-0.10 $ & $...   $ & $ -0.10$ & $ -0.06$\\
${\rm[Ca/Fe]}$       & $0.01   $ & $-0.02  $ & $...   $ & $ -0.02$ & $ -0.04$\\
${\rm[Sc/Fe]}$          & $...   $ & $-0.06   $ & $...   $ & $ -0.11$ & $ ...$\\
${\rm[Ti/Fe]}$       & $-0.05   $ & $-0.04  $ & $...   $ & $ -0.01$ & $ ...$\\
${\rm[V/Fe]}$           & $...   $ & $-0.06   $ & $...   $ & $ -0.04$ & $ ...$\\
${\rm[Cr/Fe]}$          & $0.01   $ & $0.03   $ & $...   $ & $ -0.05$ & $ ...$\\
${\rm[Co/Fe]}$          & $...   $ & $-0.06   $ & $...   $ & $ -0.05$ & $ ...$\\
${\rm[Ni/Fe]}$          & $-0.03   $ & $-0.05   $ & $...   $ & $ -0.02$ & $ -0.05$\\
${\rm[Y/Fe]}$        & $-0.09  $ & $-0.08 $ & $-0.13 $ & $ -0.11$ & $ -0.11$\\
${\rm[Zr/Fe]}$       & $...    $ & $...   $ & $-0.07 $ & $ ...$ & $ 0.02$\\
${\rm[Ba/Fe]}$       & $0.05  $ & $0.12  $ & $0.11  $ & $ 0.07$ & $ 0.06$\\
${\rm[La/Fe]}$       & $...    $ & $...   $ & $...   $ & $ ...$ & $ 0.10$\\
${\rm[Ce/Fe]}$       & $...    $ & $-0.07 $ & $-0.06 $ & $ -0.08$ & $ 0.02$\\
${\rm[Nd/Fe]}$       & $...    $ & $-0.09 $ & $...   $ & $ -0.06$ & $ -0.05$\\
${\rm[Sm/Fe]}$       & $...    $ & $...   $ & $...   $ & $ ...$ & $ 0.02$\\
${\rm[Eu/Fe]}$       & $...   $ & $-0.14 $ & $...   $ & $ -0.17$ & $ -0.06$\\
            \hline
         \end{tabular}
      \]
\end{minipage}
\tablefoot{Differences of mean [El/Fe] values for stars of Group~1 and thick-disc stars at the same metallicity interval 
$-0.57 <$ [Fe/H] $< 0.04$. 
Sixty-three stars from \citet[B14] {bensby14}, 37 stars from \citet[R06] {reddy06}, 10 stars from \citet[Ma07]{mashonkina07}, 
51 stars from \citet[S12] {stanford12}, and 10 stars from \citet[Mi13] {mishenina13}. }
   \end{table}
   
\begin{figure*}
   \centering
   \includegraphics[width=0.85\textwidth]{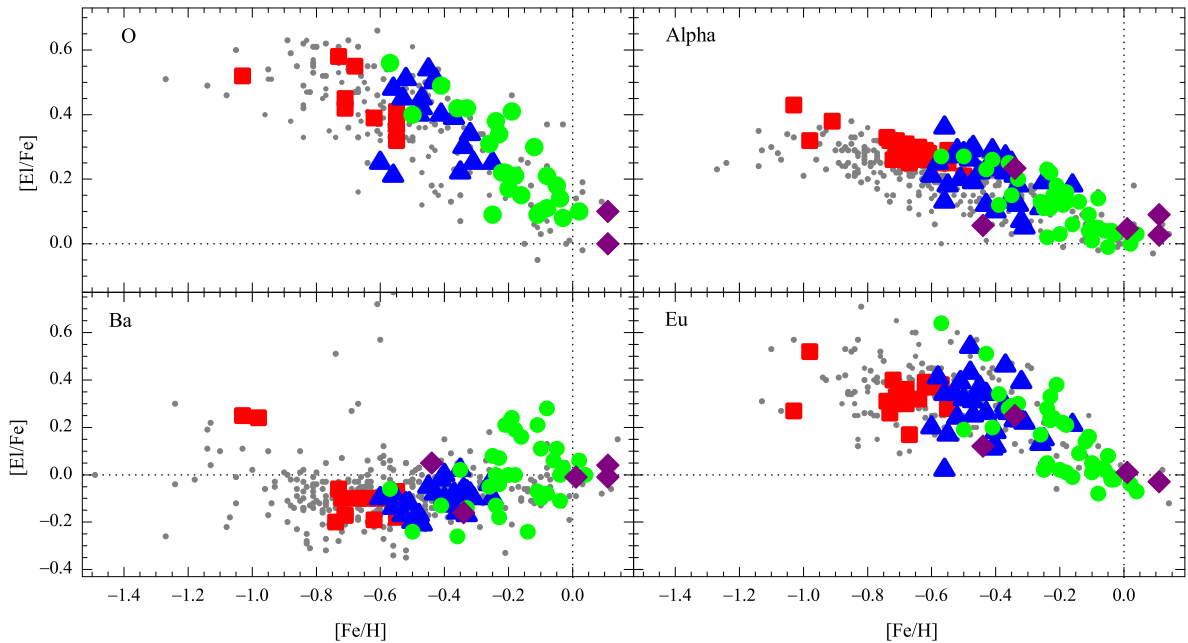}
     \caption{[El/Fe] ratio as a function of [Fe/H] in the Group 1 stars (green dots) investigated here, Group 2 
(blue triangles, Paper III), Group~3 (red squares, Papers~I and II), and comparison thick-disc stars (purple diamonds). 
    The literature data for the Milky Way thick-disc stars are shown as a grey dots.  The averaged values for $\alpha$-elements consist of Mg, Si, and Ca abundances.}
       \label{Fig.9}
  \end{figure*}

Fig.~\ref{Fig.9} displays the comparison of [El/Fe] ratios for some chemical elements between individual stars in 
Groups~1, 2, and 3 and the 
thick-disc stars of the above-mentioned studies. For comparison, we selected oxygen, the averaged values for 
the $\alpha$-elements Mg, Si, and 
Ca, and the s- and r-process-dominated elements barium and europium. Stars of the kinematic groups 
and of the thick disc have very similar chemical compositions. We have observed and analysed several thick-disc stars as well. 
Their elemental abundance ratios are 
also plotted in Fig.~\ref{Fig.9}  and agree well with results of the programme stars.     
Thus, the chemical composition of all three GCS kinematic groups is similar to the thick-disc stars, which might 
suggest that their formation histories are linked.  

\subsection{Age}

\begin{figure}
   \centering
   \includegraphics[width=0.48\textwidth]{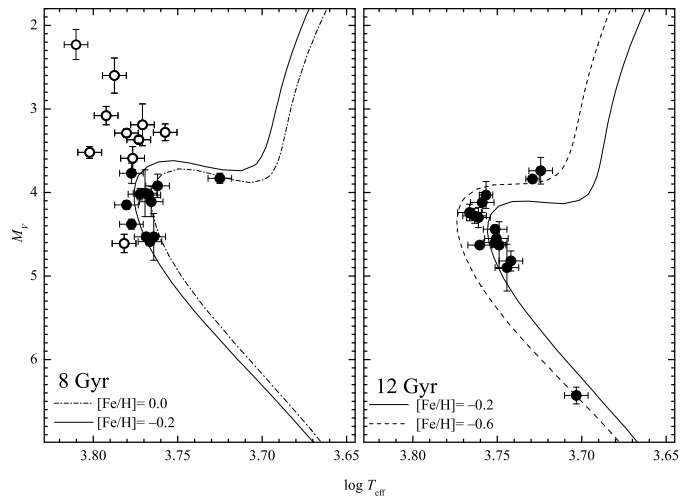}
     \caption{HR diagrams of the Group 1 stars. Isochrones are taken from \citet{bressan12}. 
     The isochrones for metal-deficient stars are with [$\alpha$/Fe ]= 0.2. The open circles 
represent the younger stars with ages of 2--5~Gyr.}
       \label{Fig.10}
   \end{figure}

According to \citet{helmi06}, the stars in Group~1 fall into two age populations: 33\% of the stars are 8~Gyr old, 
and 67\% are 12~Gyr old. 
The ages were later redetermined by \citet{holmberg09} and  \citet{casagrande11} and agree with each other within 
uncertainties (we present them in Table~1). \citet{holmberg09} present upper and lower age limits for every star. 
 The average lower and upper limit  is 10 and 14~Gyr for stars of the 12~Gyr age population,  7 and 10~Gyr for the 8~Gyr 
population,    and  4 and 5~Gyr for the younger stars, respectively.     
Fig.~\ref{Fig.10} shows the  Group 1 stars investigated here with our spectroscopic effective temperatures and absolute magnitudes $M_{v}$, taken from 
\citet{holmberg09}, in a Hertzsprung-Russell 
(HR) diagram. The isochrones were taken from  \citet{bressan12}. 
The overall features of stars in the diagrams are well reproduced by isochrones of the two indicated ages. 
The more metal-abundant stars fit the 8~Gyr isochrone quite well, while more metal-deficient stars fit the 12~Gyr 
isochrone (for metal-deficient stars, the isochrones are with [$\alpha$/Fe ]= 0.2).  A subgroup of ten stars in our sample are 
younger ($2~{\rm Gyr} \leq {\rm age} \leq 5~{\rm Gyr}$) and in the HR diagram lie higher than the turnoff 
luminosity of the 8~Gyr isochrone.  The number of stars younger than 5~Gyr among 120 stars in Group~1 is 18  (according to ages determined by \citealt{holmberg09}).  
A subgroup of about 15  young main-sequence stars can be separated among 86 Group~2 stars as well (Paper~III).  The chemical composition pattern of the young stars we investigated is similar to the rest of the GCS kinematic group stars of the same metallicity.

\subsection{Comparison with kinematic streams }

\begin{figure}
   \centering
   \includegraphics[width=0.37\textwidth]{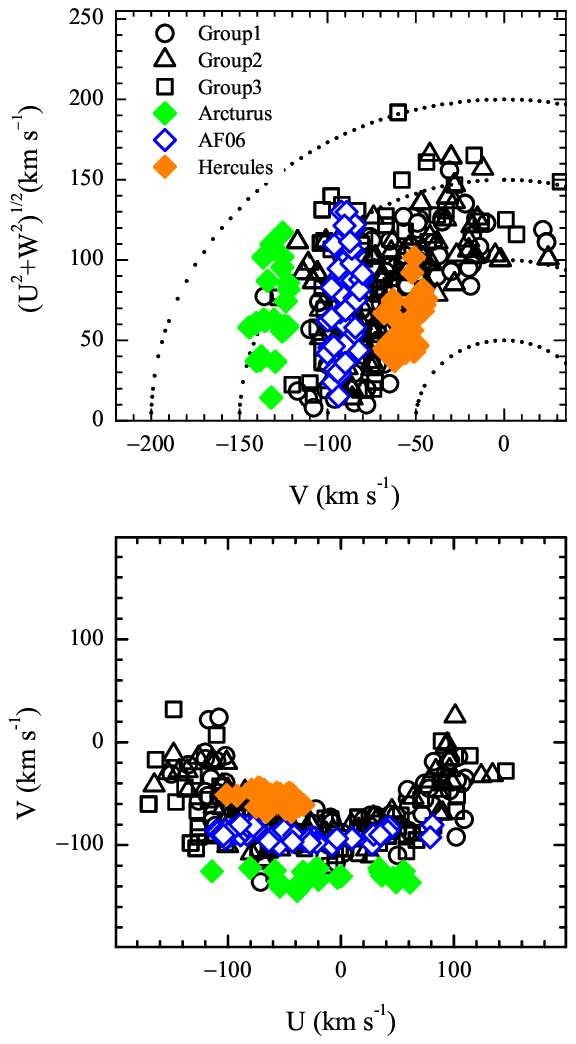}
\caption{Upper panel: Toomre diagram of stars in GCS kinematic groups and Arcturus, AF06, and Hercules streams. 
Dotted lines indicate constant values of total space 
velocity in steps of 50~km\,s$^{-1}$. Lower panel: Velocity distribution 
for stars in the same kinematic groups  and streams. } 
\label{Fig.11}
\end{figure}

In this subsection, we discuss the GCS kinematic groups in the context of three other Galactic kinematic substructures of similar metallicities.  Our attention was attracted by the Hercules stream. Stars of this kinematic stream have a  similar range of 
metallicities and ages to those in the GCS groups 
(\citealt{bobylev07, antoja08, bensby07, bensby14}). 
The Hercules stream was first identified by \citet{eggen58} as a group of 22 stars with velocities similar to the high 
velocity star $\xi $~Herculis (HD~150680).
It is believed that the Hercules stream is a result of resonant interactions between stars in the outer disc and
the Galactic bar. This stellar stream is a heterogeneous group of objects from the thin and thick discs 
(\citealt{dehnen00, fux01, quillen03, famaey05, soubiran05, pakhomov11, antoja14, bensby14}). 

The origin of the Arcturus stream has been debated for years (\citealt{eggen71}, 1996, 1998; \citealt{arifyanto06, 
gilmore02, wyse06, bensby14}, and references therein). Stars of the Arcturus stream were identified by  
Gilmore et al. (2002), and later Wyse et al. (2006), as a group of stars lagging behind the local standard of rest (LSR)
by about 100 km\,s$^{-1}$. This stream was associated with a disrupted satellite that merged with the Milky Way 10--12~Gyr ago.
 \citet{navarro04} suggested that these stars are the same group of stars that \citet{eggen71} associated with the bright 
star Arcturus, whose Galactic orbital velocity also lags at the same value.
\citet{navarro04} analysed the group of stars associated kinematically with Arcturus and confirmed that they constitute a peculiar 
grouping of metal-poor stars with a similar apocentric radius, a common angular momentum, and distinct metal abundance 
 patterns. These properties are consistent with those expected for a group of stars originating from the debris of a disrupted 
 satellite. \citet{navarro04} also noticed that the angular momentum of such a group is too low to arise from dynamical perturbations induced 
by the Galactic bar. More recently, \citet{gardner10} and \citet{monari13} showed that the Galactic long bar may produce 
a kinematic feature in velocity space with the same parameters as occupied by the Arcturus moving group. 

Another so-called AF06 stellar stream was discovered by \citet{arifyanto06} while analysing the fine structure of the 
phase space distribution function of nearby subdwarfs using data extracted 
from various catalogues. According to the discoverers,  AF06 possibly resembles the Arcturus stream. 

Fig.~\ref{Fig.11} presents a Toomre diagram and velocity distributions of stars in the GCS kinematic groups and 
Hercules, Arcturus, and AF06 streams. This diagram shows that the kinematics of stars in the GCS groups is quite 
different from the displayed streams, and only AF06 partially overlaps the pattern. 
  
\begin{figure*}
   \centering
   \includegraphics[width=0.90\textwidth]{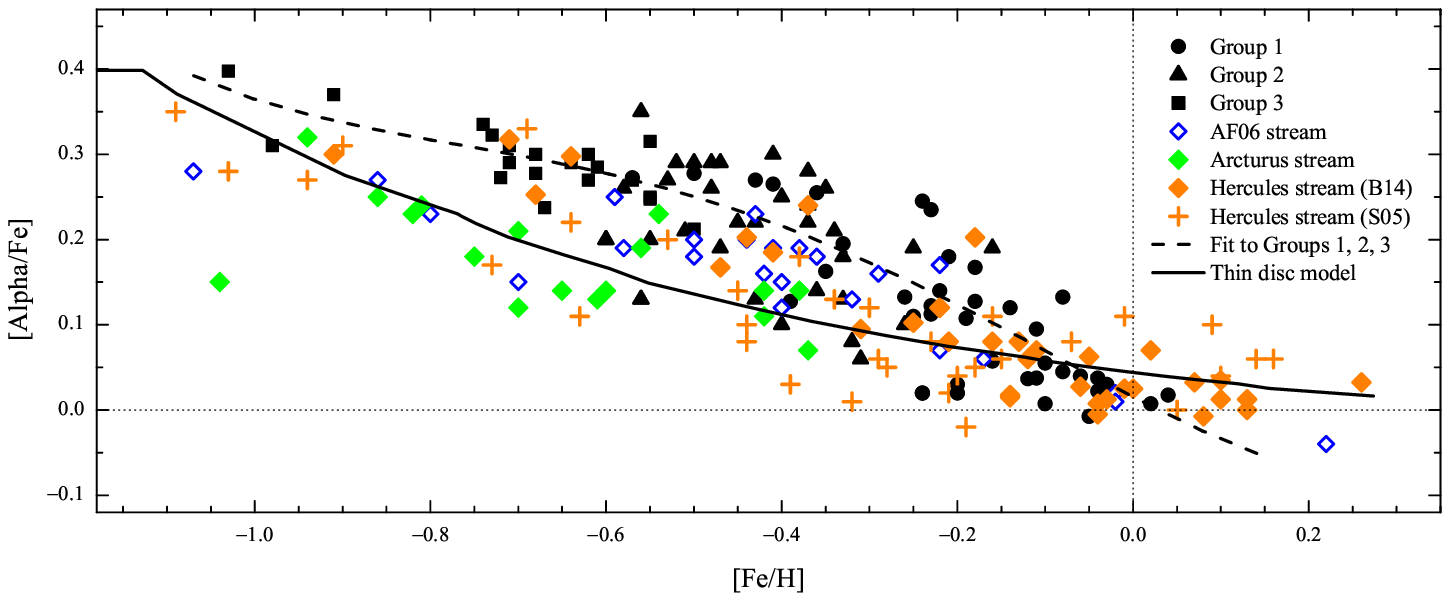}
     \caption{[$\alpha$/Fe] ratio as a function of [Fe/H] in the GCS kinematic group of stars, in the Arcturus stream 
(green diamonds), in the 
     AF06 stream (blue empty diamonds), and in the Hercules stream (orange diamonds and crosses). The data for Arcturus 
and AF06 streams were taken from \citet{ramya12}, for the Hercules stream from \citet{soubiran05} and 
     \citet{bensby14}.  The averaged values for $\alpha$-elements consist of Mg, Si, Ca, and Ti abundances. A thin-disc 
model (continuous line) is taken from \citet{pagel95}, the $2^{nd}$ 
     order polynomial fit to the GCS data is shown by the dashed line. }
       \label{Fig.12}
   \end{figure*}

A comparison of [El/Fe] ratios for $\alpha$-elements between individual stars in the GCS groups and stars of the Arcturus, 
AF06, and Hercules
streams is presented in Fig.\ref{Fig.12}. The averaged values for $\alpha$-elements consist of Mg, Si, Ca, and Ti abundances. 
We took the elemental abundances for 33 stars of the Hercules stream  from \citet{soubiran05} and for 35 stars 
from \citet{bensby14}. 
We took the elemental abundances for 18 stars of the Arcturus stream and for 26 stars of the AF06 stream  from 
\citet{ramya12}. In Fig.\ref{Fig.12} we also show the chemical evolution model of the Galactic thin-disc by \citet{pagel95} and 
a simple second-order polynomial fit to the GCS kinematic group stars.  

The element-to-iron ratios in the GCS kinematic stellar groups lie higher than in the majority of stars belonging 
to Arcturus, AF06, and Hercules streams.  The AF06 group has the chemical composition most similar to the GCS stars. 
In Paper~III, we pointed out that the Arcturus group has thick-disc kinematics, but it has seemingly  thin-disc 
abundances. Finally, as noted previously, the stars associated with the Hercules stream do not have a distinct 
chemical signature, but show a mixture of abundances as seen in the thin and thick discs (c.f. \citealt{soubiran05, bensby07, 
 bensby14, pakhomov11}). 

Thus, the presented comparison of kinematic and chemical composition patterns in the Galactic substructures, 
leads us to the conclusion that the origin of the GCS kinematic stellar groups was different from Galactic streams that\ 
originated in a course of the resonant interactions between stars in the outer disc and the Galactic bar. 

\subsection{Origin}

A sample of 274 stars was identified as an overdensity in the eccentricity range $ 0.3 < \epsilon < 0.5$ in the 
Geneva-Copenhagen survey by \citet{helmi06}. The authors provide statistical evidence that these overdensities are real,  and that they do not result from a poor choice of the comparison model of the Galaxy or uncertainties of eccentricity determinations.
It was found that stars with these dynamical characteristics do 
not constitute a homogeneous population. 
The metallicity distribution of the stars in this overdense region of the APL-space varied with eccentricity in a 
discontinuous fashion. This allowed the authors to 
separate these stars into three kinematic groups. These three groups of stars are dissimilar not only in their 
metallicity distribution, but they also have different 
kinematics in the vertical ($z$) direction.   The Group~1 velocity dispersion has $\sigma_z$  about 
$28\,{\rm km\,s}^{-1}$, that of Group~2 about 
$39\,{\rm km\,s}^{-1}$, and that of Group~3 about $52\,{\rm km\,s}^{-1}$. 

\begin{figure}
   \centering
   \includegraphics[width=0.48\textwidth]{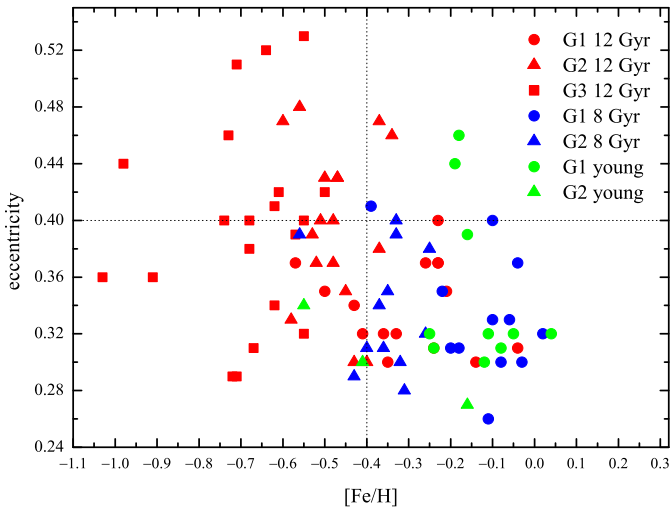}
     \caption{Eccentricity and [Fe/H] diagram of the GCS kinematic group stars. The red symbols 
     correspond to  stars investigated here with ages of about 12~Gyr, blue to about 8~Gyr, and green to the younger stars in Groups 1, 2, and 3.}
       \label{Fig.13}
   \end{figure}

\citet{helmi14}  determined the detailed chemical composition for 36 stars of the GCS Group~1, for 
22 stars of Group~2, and 14 stars of Group~3 located in the southern hemisphere. In this study, they  
noticed a relatively sharp transition in dynamical and chemical properties that occurs  at a metallicity of 
${\rm [Fe/H]}\sim -0.4$. In their sample, stars with ${\rm [Fe/H]} > -0.4$ have mostly lower eccentricities, 
smaller vertical velocity dispersions, are $\alpha$-enhanced, and define a rather narrow sequence in 
[$\alpha$/Fe] versus [Fe/H], clearly distinct from that of the thin disc. Stars with [Fe/H] $< -0.4$ have 
a range of eccentricities, are hotter vertically, and depict a larger spread in [$\alpha$/Fe].  Looking at our 
slightly larger sample of  GCS kinematic group stars (Fig.~\ref{Fig.13}) investigated here, we agree with \citet{helmi14}
that  stars with lower metallicities have a wider range of eccentricities than those that are more 
metal-abundant. However, the spread of [$\alpha$/Fe], seems to be about the 
same at all metallicities (see Fig.~\ref{Fig.9} and Fig.~\ref{Fig.12}).  In  Fig.~\ref{Fig.13}, we mark  stars of 
different ages with different colours. Practically all stars with ${\rm [Fe/H]} > -0.4$ have ages around 12~Gyr, while the youngest 
stars are predominantly metal-abundant. 

Even though, according to the dynamical characteristics, the stars of the GCS kinematic groups do not 
constitute a homogeneous population, their similar chemical composition pattern indicates that these 
kinematic groups might share a similar origin.   The similarity in chemical composition of stars in these 
kinematic groups and in the thick-disc of the Galaxy suggests 
that the formation histories of these groups and the thick-disc might be linked.  
This circumstance led us to look for the currently available thick-disc  formation scenario, which allows for the presence 
 of stars with  GCS kinematic group characteristics.  
As we pointed out in  Papers~I, II, and III, the kinematic properties of the GCS kinematic groups fit a gas-rich satellite 
merger scenario (\citealt{brook04}, 2005; \citealt{dierickx10,
wilson11, dimatteo11}) best of all.  Within this particular scenario, the eccentricities of accreted stars peak at about 
$ 0.3 < \epsilon < 0.5$ (\citealt{sales09}), which are exactly the characteristics of the  GCS groups that we investigated. 
The gas-rich merger scenario is worth  considering since, according to \citet{dierickx10}, it fits 
 the thick-disc star eccentricity distribution better than the accretion, heating, or migration scenarios. 
Although it is likely that all of those processes to some extent acted in the Milky Way, it is not clear which, if any, was the dominant mechanism. 

Finally, the numerical simulations of the disruption of a satellite galaxy that falls into its parent 
galaxy shows that the satellite debris 
can end up in several cold star streams with roughly the same characteristic eccentricities of their orbits (\citealt{helmi06}). 
The possibility of  such  a scheme   and the similar properties of element-to-iron ratios found in the GCS kinematic groups lead  to the assumption that the GCS kinematic star groups might belong to the same satellite galaxy and might have originated in our Galaxy during the same merging event.  

Investigations of formation and evolution of the Milky Way discs are continuing both observationally (e.g. \citealt{haywood13, kordopatis13b, anders14, bensby14, bergemann14, mikolaitis14}) and theoretically (\citealt{micali13, snaith14, robin14, kubryk14, minchev14}, and references therein).      
A model that could reproduce well the present day values of all of the main global observables of the 
Milky Way disc has not  been discovered yet.

\section{Conclusions}

We measured abundances of 22 chemical elements from high-resolution spectra 
in 37 stars belonging to Group~1 of the Geneva-Copenhagen survey. This kinematically identified group of stars 
as well as two other GCS kinematic groups was suggested to be a remnant of a disrupted satellite galaxy.
Our main goal was to investigate the chemical composition of stars within Group~1, to compare it with the relative 
abundance patterns in the Galactic thin- and thick-disc stars, in the GCS Group~2 and Group~3 stars, as well as in 
several kinematic streams of similar metallicities. 
         
Our study shows the following: 
  \begin{enumerate}
     \item The metallicities of the investigated stars in Group~1 are in the range of  
$0.04 \geq {\rm [Fe/H]} \geq -0.57$. The average [Fe/H] value is $-0.20\pm 0.14$~dex.

     \item Investigated Group~1 stars can be separated into three age subgroups. Along with the 
main 8- and 12-Gyr-old populations,  a subgroup of stars younger than 5~Gyr can be separated as well.  
 
    \item  All programme stars have higher abundances in oxygen, $\alpha$-elements, and r-process-dominated 
chemical elements than Galactic 
    thin-disc dwarfs and the Galactic evolution model. The abundances of iron-group chemical elements and elements 
produced mainly by the s-process are similar to those in the Galactic 
     thin-disc dwarfs of the same metallicity. 

     \item The chemical composition and kinematic properties in the GCS Group~1, 2, and 3 stars are different 
from those in stars of the Hercules, Arcturus and AF06 streams.    

    \item The chemical composition patterns in GCS Groups~1, 2, and 3 are similar to each other and to the 
thick-disc stars, which might suggest that their formation histories are linked. 
    
     \item The chemical composition together with the kinematic properties and ages of stars in the 
investigated GCS Group~1, 2 and 3 support a gas-rich satellite merger scenario as a possible 
origin for these kinematic groups.   
 
 \end{enumerate}

\begin{acknowledgements}
The data are based on observations made with the Nordic Optical Telescope, operated on the island of 
La Palma jointly by Denmark, Finland, Iceland, Norway, and Sweden, 
in the Spanish Observatorio del Roque de los Muchachos of the Instituto de Astrofisica de Canarias.  
The research leading to these results has received funding from the European Community's 
Seventh Framework Programme (FP7/2007-2013) under
grant agreement number RG226604 (OPTICON). BN acknowledges support from the Danish Research council 
and the Carlsberg Foundation. 
This research has made use of SIMBAD, VALD, and NASA ADS databases. 
\end{acknowledgements}

\Online
\appendix

{\scriptsize
\longtab{4}{
\begin{longtable}{lccccccccccccccc}
\caption{ Main atmospheric parameters and elemental abundances of  programme and comparison stars.} \label{table:4} \\

\hline\hline   
Star & $T_{\rm eff}$ & log~$g$ & $v_{t}$ & [Fe/H] & $\sigma_{\rm Fe I}$ & ${\rm n}_{\rm Fe I}$ & $\sigma_{\rm Fe II}$ & ${\rm n}_{\rm Fe II}$& [O/Fe]& [Na/Fe]& $\sigma$& n& [Mg/Fe]& $\sigma$& n\\
   & K             &       & km s$^{-1}$   &      &      &    &    &  \\
 \hline
\label{table:4}
 \endfirsthead
\caption{continued.}\\
\endhead

\hline
\noalign{\smallskip}
Group 1 stars   \\
          \noalign{\smallskip} 
 HD 3795    &   5360    & 3.7 & 1.0 &   -0.57 & 0.05 &  31 &    0.04 &  6 &       0.56    & 0.11  & 0.02 &        4 &     0.36    & 0.05  & 4 \\
  HD 4607    &  6200    & 3.8 & 1.2 &   -0.08 & 0.05 &  25 &    0.03 &  6 &       0.21    & 0.00  & 0.02 &        2 &     0.26    & 0.00  & 3 \\
  HD 15777   &  5800    & 4.2 & 1.0 &   -0.33 & 0.04 &  28 &    0.04 &  6 &       0.42    & 0.04  & 0.01 &        5 &     0.27    & 0.06  & 3 \\
  HD 22872   &  5980    & 4.0 & 1.1 &   0.04  & 0.05 &  34 &    0.02 &  8 &       ...     & -0.03 & 0.01 &        3 &     0.07    & 0.09  & 4 \\
  HD 25123   &  5880    & 3.9 & 1.1 &   0.02  & 0.04 &  37 &    0.06 &  8 &       0.10    & -0.01 & 0.01 &        5 &     0.01    & 0.02  & 3 \\
  HD 40040   &  5740    & 4.0 & 1.1 &   -0.24 & 0.05 &  37 &    0.05 &  7 &       0.38    & 0.08  & 0.02 &        3 &     0.31    & 0.06  & 4 \\
  HD 49409   &  5770    & 4.1 & 0.9 &   -0.23 & 0.05 &  32 &    0.04 &  6 &       ...     & 0.01  & 0.02 &        3 &     0.15    & 0.05  & 4 \\
  HD 52711   &  5870    & 4.1 & 1.0 &   -0.08 & 0.05 &  38 &    0.03 &  7 &       0.11    & -0.05 & 0.01 &        4 &     0.06    & 0.05  & 4 \\
  HD 60779   &  5990    & 4.1 & 0.9 &   -0.10 & 0.05 &  36 &    0.04 &  8 &       0.10    & 0.02  & 0.03 &        4 &     0.04    & 0.03  & 3 \\
  HD 67088   &  5610    & 4.0 & 0.8 &   -0.04 & 0.04 &  37 &    0.03 &  7 &       0.14    & -0.07 & 0.04 &        5 &     0.07    & 0.05  & 4 \\
  HD 67587   &  6030    & 3.8 & 1.1 &   -0.25 & 0.05 &  34 &    0.05 &  8 &       0.09    & 0.10  & 0.03 &        4 &     0.11    & 0.03  & 3 \\
  HD 76095   &  5720    & 4.1 & 1.0 &   -0.19 & 0.04 &  35 &    0.05 &  8 &       0.41    & -0.04 & 0.01 &        4 &     0.18    & 0.05  & 4 \\
  HD 77408   &  6340    & 4.2 & 1.1 &   -0.11 & 0.04 &  25 &    0.04 &  7 &       ...     & -0.04 & 0.05 &        3 &     0.21    & 0.11  & 3 \\
  HD 78558   &  5640    & 4.0 & 0.9 &   -0.41 & 0.05 &  36 &    0.04 &  7 &       0.49    & 0.12  & 0.06 &        5 &     0.33    & 0.07  & 4 \\
  HD 88371   &  5630    & 4.2 & 0.8 &   -0.23 & 0.04 &  38 &    0.06 &  8 &       0.34    & 0.01  & 0.05 &        5 &     0.29    & 0.06  & 4 \\
  HD 88446   &  5990    & 3.9 & 1.2 &   -0.39 & 0.05 &  34 &    0.03 &  8 &       ...     & 0.03  & 0.04 &        3 &     0.16    & 0.04  & 3 \\
  HD 90508   &  5760    & 4.1 & 1.0 &   -0.26 & 0.04 &  37 &    0.05 &  8 &       0.31    & -0.02 & 0.06 &        5 &     0.16    & 0.01  & 4 \\
  HD 109498  &  5810    & 4.2 & 1.0 &   -0.11 & 0.06 &  30 &    0.06 &  6 &       0.09    & 0.01  & 0.06 &        4 &     0.09    & 0.02  & 3 \\
  HD 111367  &  5830    & 4.0 & 1.0 &   -0.06 & 0.05 &  34 &    0.05 &  8 &       ...     & -0.03 & 0.01 &        4 &     0.04    & 0.05  & 3 \\
  HD 135694  &  5520    & 3.9 & 0.9 &   -0.23 & 0.06 &  24 &    0.06 &  5 &       ...     & 0.07  & 0.01 &        3 &     0.19    & 0.09  & 2 \\
  HD 138750  &  6130    & 3.8 & 1.1 &   -0.18 & 0.05 &  33 &    0.04 &  8 &       ...     & -0.02 & 0.00 &        3 &     0.24    & 0.05  & 3 \\
  HD 140209  &  5710    & 4.0 & 1.1 &   -0.14 & 0.05 &  28 &    0.03 &  5 &       ...     & 0.01  & 0.05 &        5 &     0.23    & 0.08  & 4 \\
  HD 149105  &  5930    & 3.8 & 1.0 &   -0.05 & 0.04 &  30 &    0.02 &  7 &       0.18    & -0.03 & 0.04 &        3 &     -0.03   & 0.01  & 3 \\
  HD 149890  &  6030    & 4.0 & 1.1 &   -0.20 & 0.05 &  33 &    0.03 &  6 &       0.17    & 0.00  & 0.04 &        3 &     0.07    & 0.07  & 3 \\
  HD 156617  &  5780    & 3.9 & 1.0 &   -0.03 & 0.04 &  33 &    0.03 &  7 &       0.08    & -0.09 & 0.05 &        5 &     0.07    & 0.07  & 4 \\
  HD 156893  &  5300    & 3.8 & 0.9 &   -0.21 & 0.04 &  36 &    0.03 &  8 &       0.22    & 0.02  & 0.05 &        4 &     0.23    & 0.07  & 4 \\
  HD 157214  &  5640    & 4.0 & 0.8 &   -0.36 & 0.04 &  33 &    0.02 &  6 &       0.42    & 0.07  & 0.04 &        3 &     0.32    & 0.03  & 4 \\
  BD +40 3374&  5050    & 4.6 & 0.8 &   -0.43 & 0.05 &  31 &    0.06 &  4 &       ...     & 0.05  & 0.05 &        3 &     0.27    & 0.02  & 3 \\
  HD 171009  &  5840    & 4.0 & 1.0 &   -0.35 & 0.04 &  32 &    0.04 &  6 &       ...     & -0.01 & 0.04 &        3 &     0.19    & 0.04  & 3 \\
  HD 171242  &  5920    & 3.8 & 0.9 &   -0.22 & 0.06 &  22 &    0.05 &  6 &       0.22    & 0.05  & 0.01 &        3 &     0.21    & 0.04  & 3 \\
  HD 178478  &  5550    & 3.6 & 1.0 &   -0.50 & 0.03 &  24 &    0.04 &  5 &       0.40    & 0.11  & 0.02 &        3 &     0.35    & 0.04  & 4 \\
  HD 188326  &  5310    & 3.8 & 0.9 &   -0.18 & 0.05 &  35 &    0.04 &  6 &       0.21    & 0.05  & 0.06 &        3 &     0.22    & 0.05  & 4 \\
  HD 206373  &  5900    & 3.8 & 1.1 &   -0.16 & 0.04 &  26 &    0.02 &  5 &       0.15    & 0.06  & 0.03 &        3 &     0.09    & 0.08  & 4 \\
  HD 210483  &  5850    & 4.0 & 0.9 &   -0.04 & 0.04 &  29 &    0.06 &  4 &       ...     & -0.04 & 0.05 &        5 &     0.06    & 0.10  & 4 \\
  HD 211476  &  5840    & 4.2 & 0.9 &   -0.10 & 0.05 &  32 &    0.03 &  7 &       ...     & -0.03 & 0.02 &        3 &     0.01    & 0.03  & 4 \\
  HD 217511  &  6460    & 3.7 & 1.4 &   -0.12 & 0.03 &  20 &    0.05 &  6 &       0.30    & 0.06  & 0.03 &        2 &     ...     & ...   & ... \\
  HD 219175  &  6050    & 4.2 & 0.9 &   -0.24 & 0.05 &  30 &    0.03 &  5 &       ...     & -0.04 & 0.04 &        5 &     0.05    & 0.04  & 4 \\
\noalign{\smallskip}
Thin-disc stars   \\
\noalign{\smallskip}
HD 115383 &     6090    & 4.0   & 1.2   & 0.20  & 0.05  & 29    & 0.05  & 8       & 0.02  & 0.00  & 0.01  & 4     & 0.08  & 0.07  & 3 \\  
HD 127334 &     5610    & 4.0   & 0.8   & 0.24  & 0.05  & 33    & 0.07  & 7       & -0.15 & 0.01  & 0.06  & 5     & 0.08  & 0.11  & 4 \\  
HD 136064 &     6090    & 3.9   & 1.1   & 0.05  & 0.06  & 30    & 0.04  & 7       & ...   & 0.02  & 0.03  & 3     & 0.16  & 0.14  & 2 \\  
HD 163989 &     6240    & 3.7   & 1.3   & 0.06  & 0.04  & 30    & 0.03  & 8       & 0.16  & 0.04  & 0.07  & 3     & 0.24  & 0.14  & 3 \\  
HD 187013 &     6290    & 3.6   & 1.3   & 0.00  & 0.05  & 22    & 0.04  & 7       & ...   & 0.01  & 0.04  & 3     & 0.27  & 0.01  & 2 \\  
HD 187691 &     6140    & 3.9   & 1.2   & 0.20  & 0.05  & 35    & 0.05  & 8       & 0.02  & 0.01  & 0.02  & 5     & 0.22  & 0.09  & 3 \\  
HD 200790 &     6190    & 3.9   & 1.3   & 0.08  & 0.04  & 32    & 0.02  & 8       & ...   & 0.05  & 0.05  & 3     & 0.32  & 0.09  & 2 \\  
HD 220117 &     6480    & 3.4   & 1.2   & 0.22  & 0.04  & 23    & 0.04  & 6       & ...   & ...   & ...   & ...   & ...   & ...   & ... \\
\noalign{\smallskip}
Thick-disc stars   \\
\noalign{\smallskip}
HD 150433 &     5650    & 4.2   & 0.9   & -0.34 & 0.05  & 37    & 0.05  & 7       & ...   & 0.07  & 0.01  & 3     & 0.31  & 0.04  & 4 \\
HD 181047 &     5550    & 4.2   & 0.8   & 0.01  & 0.05  & 35    & 0.04  & 6       & ...   & -0.02 & 0.02  & 5     & 0.06  & 0.05  & 3 \\
HD 186411 &     5860    & 3.7   & 1.2   & 0.11  & 0.05  & 34    & 0.03  & 8       & 0.00  & 0.00  & 0.01  & 5     & 0.16  & 0.05  & 3 \\
HD 195019 &     5750    & 4.0   & 0.9   & 0.11  & 0.03  & 37    & 0.04  & 8       & 0.10  & -0.10 & 0.01  & 5     & 0.07  & 0.08  & 4 \\
HD 198300 &     5830    & 4.1   & 0.9   & -0.44 & 0.04  & 38    & 0.04  & 8       & ...   & -0.03 & 0.03  & 3     & 0.11  & 0.05  & 4 \\
\noalign{\smallskip}
\hline
\hline
Star & [Al/Fe] & $\sigma$ & n & [Si/Fe] &$\sigma$ &  n& [Ca/Fe]&$\sigma$ &  n& [Sc/Fe]& $\sigma$& n& [Ti{\sc i}/Fe]& $\sigma$& n\\
\hline
\noalign{\smallskip}
Group 1 stars   \\
          \noalign{\smallskip}
HD 3795     & 0.38      & 0.03  & 3     & 0.20  & 0.05  & 16    & 0.25  & 0.05    & 7     & 0.16  & 0.04  & 10    & 0.28  & 0.05  & 15  \\ 
HD 4607     & 0.03      & ...   & 1     & 0.06  & 0.06  & 8     & 0.11  & 0.06    & 6     & 0.02  & 0.06  & 6     & 0.10  & 0.05  & 8   \\ 
HD 15777    & 0.20      & 0.05  & 2     & 0.18  & 0.05  & 20    & 0.15  & 0.04    & 9     & 0.22  & 0.06  & 12    & 0.18  & 0.07  & 13  \\ 
HD 22872    & -0.05     & 0.04  & 3     & 0.03  & 0.05  & 19    & -0.01 & 0.06    & 9     & 0.02  & 0.05  & 12    & -0.02 & 0.05  & 14  \\ 
HD 25123    & 0.03      & 0.01  & 2     & 0.00  & 0.04  & 20    & 0.00  & 0.05    & 11    & 0.04  & 0.04  & 10    & 0.02  & 0.05  & 18  \\ 
HD 40040    & 0.26      & 0.01  & 3     & 0.16  & 0.05  & 19    & 0.21  & 0.06    & 11    & 0.21  & 0.05  & 12    & 0.30  & 0.06  & 17  \\ 
HD 49409    & 0.04      & 0.06  & 2     & 0.05  & 0.04  & 15    & 0.13  & 0.04    & 10    & 0.10  & 0.05  & 10    & 0.16  & 0.05  & 12  \\ 
HD 52711    & -0.05     & 0.04  & 4     & 0.04  & 0.05  & 20    & 0.06  & 0.05    & 11    & -0.01 & 0.05  & 12    & 0.02  & 0.05  & 18  \\ 
HD 60779    & 0.01      & 0.04  & 2     & 0.04  & 0.05  & 19    & 0.07  & 0.05    & 8     & 0.01  & 0.05  & 11    & 0.07  & 0.05  & 15  \\ 
HD 67088    & -0.02     & 0.00  & 2     & 0.01  & 0.05  & 21    & 0.05  & 0.04    & 10    & 0.00  & 0.05  & 12    & -0.04 & 0.05  & 16  \\ 
HD 67587    & 0.11      & 0.06  & 3     & 0.09  & 0.05  & 16    & 0.14  & 0.05    & 9     & 0.05  & 0.05  & 8     & 0.10  & 0.05  & 13  \\ 
HD 76095    & 0.18      & 0.01  & 2     & 0.08  & 0.05  & 21    & 0.09  & 0.05    & 10    & 0.07  & 0.04  & 10    & 0.08  & 0.05  & 18  \\ 
HD 77408    & 0.04      & 0.04  & 3     & 0.01  & 0.06  & 14    & 0.05  & 0.04    & 7     & 0.08  & 0.05  & 9     & 0.11  & 0.05  & 8   \\ 
HD 78558    & 0.27      & 0.05  & 3     & 0.22  & 0.05  & 18    & 0.22  & 0.05    & 9     & 0.14  & 0.05  & 12    & 0.29  & 0.05  & 16  \\ 
HD 88371    & 0.25      & 0.03  & 3     & 0.16  & 0.05  & 20    & 0.22  & 0.06    & 11    & 0.22  & 0.05  & 9     & 0.27  & 0.05  & 24  \\ 
HD 88446    & 0.06      & 0.04  & 3     & 0.09  & 0.05  & 20    & 0.12  & 0.05    & 9     & 0.06  & 0.05  & 12    & 0.14  & 0.06  & 10  \\ 
HD 90508    & 0.16      & 0.04  & 3     & 0.10  & 0.05  & 21    & 0.14  & 0.04    & 11    & 0.05  & 0.04  & 11    & 0.13  & 0.05  & 18  \\ 
HD 109498   & 0.06      & 0.03  & 3     & 0.03  & 0.06  & 14    & 0.01  & 0.06    & 7     & 0.02  & 0.02  & 8     & 0.02  & 0.06  & 7   \\ 
HD 111367   & 0.07      & 0.05  & 4     & 0.03  & 0.04  & 20    & 0.04  & 0.04    & 10    & 0.04  & 0.04  & 11    & 0.05  & 0.05  & 14  \\ 
HD 135694   & 0.19      & 0.02  & 3     & 0.07  & 0.06  & 11    & 0.12  & 0.06    & 7     & 0.03  & 0.04  & 8     & 0.07  & 0.06  & 6   \\ 
HD 138750   & -0.03     & 0.03  & 2     & 0.03  & 0.05  & 20    & 0.12  & 0.05    & 9     & 0.04  & 0.04  & 11    & 0.12  & 0.05  & 10  \\ 
HD 140209   & 0.10      & ...   & 1     & 0.08  & 0.05  & 19    & 0.09  & 0.06    & 7     & 0.07  & 0.05  & 9     & 0.08  & 0.04  & 12  \\ 
HD 149105   & -0.01     & 0.03  & 2     & -0.03 & 0.04  & 15    & 0.04  & 0.07    & 5     & 0.01  & 0.04  & 8     & -0.01 & 0.05  & 7   \\ 
HD 149890   & 0.00      & 0.04  & 3     & 0.01  & 0.05  & 18    & 0.01  & 0.04    & 7     & -0.05 & 0.04  & 7     & -0.02 & 0.05  & 6   \\ 
HD 156617   & 0.06      & 0.04  & 4     & -0.01 & 0.04  & 18    & 0.03  & 0.05    & 11    & -0.02 & 0.04  & 10    & 0.03  & 0.05  & 15  \\ 
HD 156893   & 0.29      & 0.04  & 3     & 0.14  & 0.04  & 20    & 0.16  & 0.05    & 11    & 0.18  & 0.05  & 11    & 0.19  & 0.05  & 25  \\ 
HD 157214   & 0.31      & 0.05  & 2     & 0.22  & 0.04  & 16    & 0.22  & 0.06    & 8     & 0.15  & 0.04  & 9     & 0.26  & 0.04  & 12  \\ 
BD +40 3374 & 0.38      & 0.03  & 3     & 0.20  & 0.05  & 18    & 0.23  & 0.04    & 8     & 0.24  & 0.04  & 8     & 0.38  & 0.07  & 22  \\ 
HD 171009   & 0.12      & ...   & 1     & 0.12  & 0.04  & 15    & 0.15  & 0.06    & 7     & 0.13  & 0.04  & 7     & 0.19  & 0.03  & 8   \\ 
HD 171242   & 0.01      & 0.04  & 2     & 0.10  & 0.05  & 11    & 0.14  & 0.04    & 7     & 0.02  & 0.05  & 7     & 0.11  & 0.04  & 10  \\ 
HD 178478   & 0.33      & 0.04  & 3     & 0.20  & 0.05  & 14    & 0.25  & 0.05    & 7     & 0.06  & 0.06  & 7     & 0.31  & 0.06  & 8   \\ 
HD 188326   & 0.22      & 0.00  & 2     & 0.13  & 0.03  & 15    & 0.12  & 0.05    & 7     & 0.07  & 0.05  & 6     & 0.20  & 0.05  & 13  \\ 
HD 206373   & 0.02      & 0.03  & 4     & 0.04  & 0.05  & 15    & 0.05  & 0.02    & 6     & 0.04  & 0.04  & 8     & 0.05  & 0.03  & 7   \\ 
HD 210483   & 0.07      & 0.03  & 3     & 0.00  & 0.05  & 17    & 0.05  & 0.06    & 7     & 0.02  & 0.04  & 8     & 0.04  & 0.05  & 18  \\ 
HD 211476   & 0.05      & 0.01  & 2     & -0.01 & 0.04  & 16    & 0.02  & 0.03    & 6     & -0.03 & 0.04  & 8     & 0.01  & 0.03  & 9   \\ 
HD 217511   & 0.00      & 0.04  & 2     & 0.03  & 0.05  & 17    & 0.06  & 0.05    & 9     & 0.10  & 0.05  & 7     & 0.02  & 0.05  & 6   \\ 
HD 219175   & -0.06     & ...   & 1     & -0.01 & 0.06  & 15    & 0.01  & 0.05    & 7     & -0.02 & 0.03  & 9     & 0.02  & 0.04  & 6   \\            
\noalign{\smallskip}
Thin-disc stars \\
\noalign{\smallskip}
HD 115383     & ...     & ...   & ...   & 0.06  & 0.05  & 13    & 0.08  & 0.04    & 8     & 0.00  & 0.06  & 9     & 0.07  & 0.04  & 8   \\
HD 127334     & 0.07    & 0.03  & 2     & 0.03  & 0.05  & 20    & 0.00  & 0.05    & 11    & -0.02 & 0.05  & 10    & -0.03 & 0.05  & 16  \\
HD 136064     & -0.04   & 0.03  & 2     & 0.00  & 0.05  & 15    & 0.09  & 0.05    & 6     & 0.10  & 0.04  & 8     & -0.02 & 0.05  & 10  \\
HD 163989     & 0.04    & 0.07  & 4     & 0.06  & 0.05  & 15    & 0.04  & 0.03    & 4     & 0.06  & 0.05  & 9     & 0.10  & 0.06  & 8   \\
HD 187013     & -0.04   & 0.02  & 2     & 0.06  & 0.06  & 11    & 0.09  & 0.04    & 4     & 0.01  & 0.06  & 7     & 0.09  & 0.06  & 5   \\
HD 187691     & 0.00    & 0.06  & 3     & 0.03  & 0.05  & 18    & 0.05  & 0.02    & 8     & 0.01  & 0.05  & 12    & 0.03  & 0.05  & 12  \\
HD 200790     & 0.01    & 0.01  & 2     & 0.04  & 0.05  & 16    & 0.01  & 0.04    & 7     & 0.05  & 0.05  & 9     & 0.00  & 0.04  & 7   \\
HD 220117     & 0.01    & ...   & 1     & 0.00  & 0.05  & 9     & 0.10  & 0.05    & 5     & 0.06  & 0.04  & 4     & 0.12  & 0.05  & 6   \\
\noalign{\smallskip}
Thick-disc stars \\
\noalign{\smallskip}
HD 150433     & 0.10    & ...   & 1     & 0.17  & 0.05  & 18    & 0.22  & 0.05    & 9     & 0.19  & 0.05  & 9     & 0.28  & 0.05  & 18  \\
HD 181047     & 0.05    & 0.02  & 2     & 0.03  & 0.04  & 20    & 0.05  & 0.05    & 11    & -0.02 & 0.05  & 11    & -0.01 & 0.05  & 23  \\
HD 186411     & 0.05    & 0.02  & 3     & 0.00  & 0.05  & 20    & 0.11  & 0.04    & 8     & 0.02  & 0.04  & 9     & 0.03  & 0.05  & 18  \\
HD 195019     & -0.02   & 0.03  & 3     & -0.01 & 0.04  & 20    & 0.02  & 0.05    & 10    & -0.04 & 0.05  & 11    & -0.02 & 0.05  & 25  \\
HD 198300     & 0.18    & 0.05  & 4     & 0.03  & 0.03  & 19    & 0.03  & 0.05    & 9     & -0.01 & 0.05  & 11    & 0.1   & 0.06  & 11  \\
\noalign{\smallskip}
\hline
\hline
Star & [Ti{\sc ii}/Fe] & $\sigma$ & n & [V/Fe] & $\sigma$ & n & [Cr/Fe] &$\sigma$ &  n& [Co/Fe]& $\sigma$& n& [Ni/Fe]& $\sigma$& n\\
\hline
\noalign{\smallskip}
Group 1 stars   \\
          \noalign{\smallskip}
HD 3795     & 0.20      & 0.06  & 4     & 0.15  & 0.05  & 11    & 0.07  & 0.05    & 13    & 0.15  & 0.03  & 6     & 0.04  & 0.05  & 23  \\
HD 4607     & 0.06      & 0.02  & 3     & 0.03  & 0.03  & 5     & -0.01 & 0.06    & 13    & -0.06 & 0.05  & 5     & -0.04 & 0.05  & 15  \\
HD 15777    & 0.22      & 0.03  & 4     & 0.12  & 0.05  & 8     & 0.02  & 0.05    & 17    & 0.12  & 0.05  & 8     & 0.00  & 0.05  & 27  \\
HD 22872    & -0.03     & 0.06  & 4     & 0.02  & 0.05  & 13    & 0.00  & 0.05    & 19    & -0.05 & 0.05  & 10    & -0.04 & 0.05  & 30  \\
HD 25123    & 0.02      & 0.04  & 4     & 0.00  & 0.04  & 10    & 0.00  & 0.06    & 17    & 0.00  & 0.05  & 9     & -0.02 & 0.06  & 30  \\
HD 40040    & 0.26      & 0.04  & 4     & 0.14  & 0.06  & 12    & 0.03  & 0.06    & 17    & 0.11  & 0.05  & 6     & 0.04  & 0.04  & 28  \\
HD 49409    & 0.08      & 0.02  & 4     & 0.14  & 0.05  & 10    & 0.02  & 0.05    & 14    & 0.03  & 0.03  & 7     & -0.01 & 0.05  & 19  \\
HD 52711    & 0.03      & 0.06  & 4     & -0.02 & 0.05  & 12    & 0.01  & 0.05    & 18    & -0.02 & 0.05  & 10    & -0.02 & 0.05  & 29  \\
HD 60779    & 0.03      & 0.05  & 4     & 0.06  & 0.06  & 9     & 0.01  & 0.06    & 15    & -0.03 & 0.06  & 9     & -0.02 & 0.05  & 28  \\
HD 67088    & -0.05     & 0.06  & 4     & -0.06 & 0.06  & 14    & 0.00  & 0.05    & 19    & -0.07 & 0.04  & 9     & -0.05 & 0.05  & 29  \\
HD 67587    & 0.06      & 0.02  & 4     & 0.13  & 0.05  & 6     & 0.03  & 0.06    & 14    & 0.06  & 0.04  & 5     & -0.03 & 0.05  & 23  \\
HD 76095    & 0.12      & 0.04  & 3     & 0.06  & 0.05  & 12    & 0.05  & 0.05    & 17    & 0.05  & 0.04  & 9     & -0.02 & 0.05  & 30  \\
HD 77408    & 0.10      & 0.07  & 3     & 0.07  & 0.05  & 5     & -0.02 & 0.05    & 11    & 0.03  & 0.04  & 6     & -0.05 & 0.05  & 14  \\
HD 78558    & 0.20      & 0.05  & 4     & 0.07  & 0.05  & 12    & 0.04  & 0.06    & 18    & 0.08  & 0.04  & 8     & 0.02  & 0.05  & 28  \\
HD 88371    & 0.27      & 0.02  & 4     & 0.13  & 0.04  & 14    & 0.04  & 0.06    & 18    & 0.07  & 0.03  & 10    & 0.04  & 0.05  & 28  \\
HD 88446    & 0.04      & 0.02  & 4     & 0.02  & 0.03  & 5     & 0.00  & 0.05    & 14    & 0.04  & 0.02  & 5     & -0.05 & 0.04  & 23  \\
HD 90508    & 0.07      & 0.04  & 4     & 0.06  & 0.05  & 13    & 0.01  & 0.05    & 18    & 0.03  & 0.05  & 10    & -0.02 & 0.05  & 30  \\
HD 109498   & 0.01      & 0.03  & 4     & 0.00  & 0.04  & 6     & 0.01  & 0.07    & 15    & 0.01  & 0.03  & 4     & -0.04 & 0.06  & 18  \\
HD 111367   & -0.01     & 0.04  & 4     & 0.01  & 0.06  & 7     & -0.01 & 0.04    & 16    & -0.02 & 0.05  & 8     & -0.04 & 0.05  & 28  \\
HD 135694   & 0.05      & 0.00  & 2     & -0.03 & 0.05  & 9     & 0.02  & 0.04    & 12    & 0.01  & 0.03  & 6     & -0.07 & 0.05  & 14  \\
HD 138750   & 0.08      & 0.08  & 4     & 0.04  & 0.05  & 5     & 0.00  & 0.05    & 14    & -0.02 & 0.05  & 6     & -0.06 & 0.05  & 24  \\
HD 140209   & 0.04      & 0.03  & 3     & 0.04  & 0.05  & 9     & 0.04  & 0.05    & 15    & 0.01  & 0.05  & 7     & -0.03 & 0.06  & 26  \\
HD 149105   & -0.06     & 0.03  & 2     & -0.08 & 0.01  & 3     & -0.01 & 0.09    & 15    & -0.08 & 0.06  & 5     & -0.04 & 0.03  & 24  \\
HD 149890   & -0.01     & 0.02  & 3     & -0.02 & 0.04  & 5     & -0.02 & 0.07    & 13    & -0.06 & 0.04  & 9     & -0.07 & 0.05  & 21  \\
HD 156617   & -0.04     & 0.05  & 4     & -0.05 & 0.04  & 12    & 0.02  & 0.05    & 17    & -0.05 & 0.03  & 9     & -0.07 & 0.05  & 29  \\
HD 156893   & 0.17      & 0.05  & 4     & 0.08  & 0.05  & 14    & 0.02  & 0.06    & 16    & 0.05  & 0.04  & 11    & 0.00  & 0.05  & 27  \\
HD 157214   & 0.10      & 0.01  & 2     & 0.13  & 0.03  & 5     & 0.02  & 0.08    & 12    & 0.16  & 0.03  & 5     & 0.03  & 0.06  & 19  \\
BD +40 3374 & 0.30      & 0.04  & 3     & 0.24  & 0.04  & 7     & 0.10  & 0.05    & 13    & 0.19  & 0.04  & 8     & 0.04  & 0.06  & 24  \\
HD 171009   & 0.11      & 0.04  & 3     & 0.07  & 0.04  & 4     & -0.02 & 0.08    & 13    & 0.06  & 0.08  & 5     & -0.03 & 0.05  & 16  \\
HD 171242   & 0.03      & 0.05  & 4     & 0.03  & 0.04  & 8     & 0.03  & 0.04    & 12    & 0.05  & 0.05  & 4     & -0.09 & 0.05  & 15  \\
HD 178478   & 0.21      & 0.01  & 3     & 0.16  & 0.05  & 8     & 0.10  & 0.06    & 9     & 0.07  & 0.03  & 4     & -0.03 & 0.05  & 18  \\
HD 188326   & 0.27      & 0.03  & 4     & 0.14  & 0.06  & 9     & 0.05  & 0.05    & 15    & 0.10  & 0.04  & 7     & 0.02  & 0.05  & 21  \\
HD 206373   & 0.03      & 0.05  & 4     & 0.01  & 0.05  & 5     & -0.04 & 0.05    & 13    & -0.03 & 0.04  & 4     & -0.04 & 0.04  & 18  \\
HD 210483   & 0.01      & 0.04  & 4     & -0.06 & 0.02  & 7     & 0.01  & 0.07    & 14    & -0.04 & 0.05  & 9     & -0.06 & 0.05  & 26  \\
HD 211476   & 0.00      & 0.04  & 4     & -0.03 & 0.04  & 6     & 0.01  & 0.05    & 16    & -0.04 & 0.04  & 7     & -0.04 & 0.04  & 22  \\
HD 217511   & 0.06      & 0.01  & 2     & ...   & ...   & ...   & 0.00  & 0.05    & 8     & 0.04  & 0.01  & 4     & -0.01 & 0.05  & 13  \\
HD 219175   & -0.03     & 0.03  & 3     & -0.02 & 0.05  & 8     & -0.05 & 0.05    & 9     & -0.04 & 0.03  & 3     & -0.08 & 0.05  & 22  \\
\noalign{\smallskip}
Thin-disc stars \\
\noalign{\smallskip}
HD 115383     & 0.07    & 0.03  & 3     & 0.00  & 0.06  & 7     & 0.02  & 0.06    & 16    & 0.02  & 0.05  & 5     & -0.01 & 0.04  & 20  \\
HD 127334     & -0.07   & 0.04  & 4     & -0.03 & 0.05  & 13    & 0.03  & 0.05    & 15    & 0.02  & 0.04  & 8     & 0.02  & 0.05  & 26  \\
HD 136064     & 0.05    & 0.03  & 3     & 0.01  & 0.05  & 8     & 0.03  & 0.05    & 12    & -0.04 & 0.03  & 6     & -0.04 & 0.05  & 16  \\
HD 163989     & 0.08    & 0.03  & 4     & 0.02  & 0.08  & 4     & -0.03 & 0.05    & 11    & -0.03 & 0.05  & 6     & -0.07 & 0.05  & 23  \\
HD 187013     & 0.09    & 0.01  & 2     & ...   & ...   & ...   & 0.02  & 0.06    & 9     & ...   & ...   & ...   & -0.04 & 0.05  & 13  \\
HD 187691     & 0.02    & 0.03  & 4     & -0.04 & 0.05  & 7     & 0.00  & 0.04    & 17    & -0.07 & 0.06  & 6     & -0.03 & 0.05  & 28  \\
HD 200790     & -0.04   & 0.03  & 4     & -0.04 & 0.05  & 5     & 0.01  & 0.04    & 15    & -0.04 & 0.05  & 6     & -0.03 & 0.04  & 23  \\
HD 220117     & ...     & ...   & ...   & ...   & ...   & ...   & 0.00  & 0.06    & 7     & ...   & ...   & ...   & -0.04 & 0.04  & 9   \\
\noalign{\smallskip}
Thick-disc stars \\
\noalign{\smallskip}
HD 150433     & 0.26    & 0.03  & 2     & 0.17  & 0.05  & 8     & 0.07  & 0.05    & 16    & 0.07  & 0.05  & 7     & 0.04  & 0.05  & 25  \\
HD 181047     & -0.05   & 0.05  & 4     & -0.02 & 0.05  & 14    & 0.05  & 0.06    & 18    & -0.02 & 0.06  & 11    & -0.01 & 0.05  & 26  \\
HD 186411     & 0.02    & 0.03  & 4     & -0.01 & 0.06  & 12    & 0.01  & 0.06    & 17    & -0.04 & 0.05  & 7     & -0.07 & 0.05  & 27  \\
HD 195019     & -0.05   & 0.05  & 4     & -0.05 & 0.05  & 12    & 0.01  & 0.05    & 19    & -0.05 & 0.05  & 8     & -0.06 & 0.05  & 30  \\
HD 198300     & 0.03    & 0.05  & 4     & 0.01  & 0.04  & 7     & -0.04 & 0.06    & 16    & 0.04  & 0.02  & 6     & -0.06 & 0.05  & 26  \\
\hline
\end{longtable}
}

\longtab{5}{
\begin{longtable}{lccccccccccccccc}
\caption{Elemental abundances of neutron-capture elements for programme and comparison stars.} \\
\hline\hline 
Star & [Y/Fe] & $\sigma$ & n & [Zr{\sc i}/Fe] & $\sigma$ & n & [Zr{\sc ii}/Fe] & $\sigma$ & n &[Ba/Fe] & [La/Fe]& $\sigma$& n \\
\hline
\label{table:5}
\endfirsthead
\caption{continued.}\\
\endhead
\hline
\noalign{\smallskip}
Group 1 stars   \\
          \noalign{\smallskip} 
HD 3795     & -0.03     & 0.03  & 6     & 0.17  & 0.07  & 4     & 0.08  & 0.02    & 2     & -0.06 & 0.20  & 0.05  & 3  \\ 
HD 4607     & -0.02     & 0.02  & 5     & 0.07  & 0.02  & 3     & 0.07  & 0.01    & 2     & 0.28  & -0.05 & 0.07  & 2   \\
HD 15777    & -0.13     & 0.06  & 4     & 0.13  & 0.00  & 3     & 0.10  & 0.09    & 2     & -0.14 & -0.01 & 0.01  & 3   \\
HD 22872    & -0.12     & 0.01  & 6     & -0.06 & 0.08  & 4     & -0.01 & 0.09    & 2     & 0.00  & -0.16 & 0.05  & 3   \\
HD 25123    & -0.07     & 0.05  & 7     & 0.07  & 0.02  & 2     & -0.04 & 0.01    & 2     & 0.06  & 0.00  & 0.04  & 4   \\
HD 40040    & -0.22     & 0.03  & 4     & 0.03  & 0.01  & 2     & -0.05 & 0.05    & 2     & -0.13 & 0.02  & 0.06  & 2   \\
HD 49409    & -0.21     & 0.08  & 2     & 0.19  & 0.05  & 2     & 0.11  & ...     & 1     & 0.07  & -0.13 & ...   & 1   \\
HD 52711    & -0.16     & 0.03  & 7     & 0.08  & 0.05  & 5     & 0.05  & 0.04    & 2     & -0.08 & -0.06 & 0.06  & 4   \\
HD 60779    & -0.06     & 0.06  & 7     & 0.05  & 0.03  & 3     & 0.01  & 0.04    & 2     & 0.11  & -0.04 & 0.06  & 4   \\
HD 67088    & -0.20     & 0.04  & 7     & -0.20 & 0.03  & 3     & -0.15 & 0.04    & 2     & 0.00  & -0.03 & 0.05  & 4   \\
HD 67587    & -0.16     & 0.05  & 4     & -0.10 & ...   & 1     & -0.18 & ...     & 1     & 0.08  & 0.01  & 0.00  & 3   \\
HD 76095    & -0.05     & 0.06  & 5     & 0.00  & 0.05  & 3     & 0.00  & 0.04    & 2     & 0.24  & 0.14  & 0.01  & 2   \\
HD 77408    & -0.08     & 0.01  & 3     & ...   & ...   & ...   & 0.17  & ...     & 1     & 0.21  & 0.07  & ...   & 1   \\
HD 78558    & -0.13     & 0.05  & 6     & 0.07  & 0.09  & 2     & -0.01 & 0.01    & 2     & -0.13 & -0.16 & 0.04  & 2   \\
HD 88371    & 0.03      & 0.07  & 4     & 0.06  & 0.03  & 4     & 0.06  & 0.02    & 2     & -0.18 & -0.09 & ...   & 1   \\
HD 88446    & 0.57      & 0.05  & 5     & 0.57  & ...   & 1     & 0.56  & 0.01    & 2     & 1.04  & 0.70  & 0.08  & 4   \\
HD 90508    & -0.19     & 0.05  & 6     & -0.06 & 0.05  & 4     & -0.01 & 0.01    & 2     & -0.05 & 0.02  & 0.04  & 3   \\
HD 109498   & -0.09     & 0.02  & 6     & -0.02 & 0.06  & 3     & -0.03 & 0.04    & 2     & -0.07 & -0.01 & 0.07  & 2   \\
HD 111367   & -0.17     & 0.05  & 6     & 0.00  & 0.06  & 3     & -0.02 & 0.03    & 2     & 0.06  & 0.06  & 0.05  & 2   \\
HD 135694   & -0.20     & 0.07  & 3     & -0.01 & 0.07  & 4     & 0.01  & ...     & 1     & -0.04 & ...   & ...   & ... \\
HD 138750   & 0.00      & 0.04  & 7     & 0.16  & 0.03  & 3     & 0.04  & 0.07    & 2     & 0.19  & 0.04  & 0.05  & 4   \\
HD 140209   & -0.13     & 0.03  & 4     & 0.00  & 0.04  & 3     & -0.14 & 0.06    & 2     & -0.24 & -0.07 & 0.02  & 3   \\
HD 149105   & -0.16     & 0.04  & 5     & -0.02 & ...   & 1     & -0.03 & 0.01    & 2     & 0.11  & -0.07 & 0.00  & 2   \\
HD 149890   & -0.15     & 0.04  & 6     & 0.09  & ...   & 1     & 0.07  & ...     & 1     & 0.00  & -0.02 & 0.01  & 2   \\
HD 156617   & -0.21     & 0.05  & 6     & -0.09 & 0.04  & 2     & -0.07 & ...     & 1     & 0.03  & -0.07 & 0.08  & 3   \\
HD 156893   & -0.03     & 0.06  & 6     & -0.02 & 0.04  & 6     & 0.04  & 0.02    & 2     & 0.21  & 0.15  & 0.07  & 3   \\
HD 157214   & -0.13     & 0.07  & 5     & 0.08  & 0.06  & 4     & 0.05  & 0.03    & 2     & -0.26 & -0.06 & 0.06  & 2   \\
BD +40 3374 & -0.14     & 0.07  & 3     & 0.30  & 0.04  & 5     & 0.34  & ...     & 1     & ...   & 0.04  & ...   & 1   \\
HD 171009   & -0.12     & 0.04  & 4     & 0.17  & 0.03  & 3     & 0.09  & ...     & 1     & 0.02  & 0.05  & 0.05  & 3   \\
HD 171242   & -0.14     & 0.00  & 2     & 0.24  & ...   & 1     & 0.21  & 0.06    & 2     & ...   & 0.20  & 0.01  & 2   \\
HD 178478   & -0.15     & 0.10  & 3     & 0.18  & 0.04  & 2     & ...   & ...     & ...   & -0.24 & 0.11  & 0.05  & 3   \\
HD 188326   & -0.05     & 0.06  & 5     & 0.06  & 0.03  & 6     & 0.02  & ...     & 1     & 0.00  & 0.13  & 0.05  & 3   \\
HD 206373   & -0.09     & 0.06  & 6     & ...   & ...   & ...   & -0.14 & ...     & 1     & 0.16  & -0.07 & 0.05  & 3   \\
HD 210483   & -0.17     & 0.05  & 4     & 0.00  & 0.02  & 2     & -0.05 & 0.02    & 2     & -0.11 & -0.03 & 0.06  & 3   \\
HD 211476   & -0.16     & 0.06  & 7     & -0.04 & 0.04  & 5     & 0.00  & 0.06    & 2     & -0.10 & 0.06  & 0.05  & 2   \\
HD 217511   & 0.11      & 0.02  & 5     & ...   & ...   & ...   & ...   & ...     & ...   & ...   & 0.21  & 0.01  & 2   \\
HD 219175   & -0.15     & 0.04  & 5     & -0.09 & ...   & 1     & -0.09 & 0.00    & 2     & 0.00  & 0.12  & 0.00  & 2   \\
\noalign{\smallskip}
Thin-disc stars \\
\noalign{\smallskip}
HD 115383     & -0.17   & 0.06  & 5     & 0.00  & ...   & 1     & -0.03 & 0.05    & 2     & -0.05 & -0.09 & ...   & 1   \\
HD 127334     & -0.20   & 0.05  & 6     & -0.15 & 0.05  & 7     & -0.15 & 0.03    & 2     & ...   & -0.17 & 0.06  & 3   \\
HD 136064     & 0.05    & 0.00  & 2     & ...   & ...   & ...   & 0.19  & 0.04    & 2     & 0.05  & 0.08  & ...   & 1   \\
HD 163989     & 0.01    & 0.05  & 5     & ...   & ...   & ...   & 0.03  & 0.01    & 2     & 0.03  & 0.08  & 0.09  & 3   \\
HD 187013     & -0.15   & 0.03  & 5     & -0.12 & ...   & 1     & -0.12 & ...     & 1     & 0.00  & 0.07  & 0.04  & 2   \\
HD 187691     & -0.13   & 0.07  & 6     & -0.10 & ...   & 1     & -0.09 & 0.01    & 2     & -0.14 & -0.05 & 0.02  & 3   \\
HD 200790     & -0.18   & 0.04  & 5     & -0.12 & ...   & 1     & -0.11 & 0.07    & 2     & -0.17 & -0.07 & 0.05  & 4   \\
HD 220117     & ...     & ...   & ...   & ...   & ...   & ...   & ...   & ...     & ...   & -0.17 & ...   & ...   & ... \\
\noalign{\smallskip}
Thick-disc stars \\
\noalign{\smallskip}
HD 150433     & -0.06   & 0.04  & 5     & 0.12  & 0.06  & 4     & 0.12  & ...     & 1     & -0.16 & 0.00  & 0.04  & 2   \\ 
HD 181047     & 0.00    & 0.06  & 7     & -0.08 & 0.05  & 4     & -0.05 & 0.01    & 2     & -0.01 & -0.06 & 0.02  & 3   \\ 
HD 186411     & -0.05   & 0.05  & 6     & -0.03 & 0.04  & 3     & -0.03 & 0.04    & 2     & 0.04  & 0.03  & 0.06  & 4   \\ 
HD 195019     & -0.17   & 0.06  & 7     & -0.20 & 0.04  & 4     & -0.17 & 0.04    & 2     & -0.01 & -0.09 & 0.06  & 4   \\ 
HD 198300     & -0.14   & 0.06  & 5     & -0.02 & 0.09  & 2     & ...   & ...     & ...   & 0.05  & -0.01 & 0.07  & 3   \\ 
\noalign{\smallskip}  
\hline\hline 
Star & [Ce/Fe] & $\sigma$ & n & [Pr/Fe] & $\sigma$ & n & [Nd/Fe] &  $\sigma$ & n &  [Sm/Fe]  &  $\sigma$ & n &  [Eu/Fe]  &  $\sigma$ & n \\
\hline
\noalign{\smallskip}
Group 1 stars   \\
          \noalign{\smallskip}
HD 3795     & 0.17      & 0.06  & 2     & 0.44  & 0.14  & 2     & 0.32  & 0.03    & 6     & 0.44  & 0.13  & 3     & 0.64  & 0.03  & 2     \\  
HD 4607     & -0.02     & 0.00  & 2     & 0.14  & ...   & 1     & 0.08  & 0.04    & 4     & 0.11  & 0.01  & 2     & 0.03  & 0.04  & 2      \\ 
HD 15777    & 0.04      & 0.02  & 3     & 0.14  & 0.05  & 2     & 0.12  & 0.05    & 5     & 0.18  & 0.05  & 3     & 0.30  & 0.04  & 2      \\ 
HD 22872    & -0.09     & 0.09  & 3     & 0.02  & 0.06  & 2     & -0.10 & 0.02    & 7     & -0.01 & 0.08  & 3     & -0.07 & 0.06  & 2      \\ 
HD 25123    & -0.08     & 0.05  & 5     & -0.05 & 0.01  & 2     & -0.06 & 0.05    & 5     & -0.06 & 0.04  & 3     & -0.04 & 0.00  & 2      \\ 
HD 40040    & -0.05     & 0.06  & 3     & 0.24  & 0.02  & 2     & 0.09  & 0.05    & 5     & 0.21  & 0.05  & 2     & 0.28  & 0.03  & 2      \\ 
HD 49409    & -0.04     & 0.06  & 2     & 0.23  & ...   & 1     & 0.25  & 0.03    & 3     & 0.24  & 0.07  & 3     & 0.33  & ...   & 1      \\ 
HD 52711    & -0.09     & 0.05  & 4     & -0.01 & 0.12  & 2     & -0.04 & 0.05    & 5     & -0.01 & 0.03  & 3     & -0.08 & 0.04  & 2      \\ 
HD 60779    & 0.03      & 0.04  & 3     & 0.19  & ...   & 1     & -0.02 & 0.07    & 6     & 0.01  & 0.03  & 3     & 0.01  & 0.02  & 2      \\ 
HD 67088    & -0.10     & 0.03  & 4     & 0.06  & 0.03  & 2     & -0.08 & 0.05    & 6     & -0.03 & 0.06  & 3     & -0.01 & 0.08  & 2      \\ 
HD 67587    & -0.06     & 0.05  & 2     & ...   & ...   & ...   & 0.07  & 0.07    & 3     & ...   & ...   & ...   & 0.02  & 0.06  & 2     \\  
HD 76095    & 0.13      & 0.04  & 4     & 0.16  & 0.06  & 2     & 0.16  & 0.07    & 5     & 0.15  & 0.01  & 2     & 0.22  & 0.04  & 2      \\ 
HD 77408    & 0.06      & ...   & 1     & 0.21  & ...   & 1     & 0.20  & 0.02    & 3     & 0.22  & ...   & 1     & 0.16  & 0.02  & 2      \\ 
HD 78558    & -0.02     & 0.05  & 5     & 0.22  & 0.03  & 2     & 0.03  & 0.05    & 5     & 0.16  & 0.01  & 2     & 0.20  & 0.01  & 2      \\ 
HD 88371    & 0.00      & 0.04  & 3     & 0.11  & ...   & 1     & 0.08  & 0.03    & 5     & 0.21  & 0.01  & 3     & 0.24  & ...   & 1      \\ 
HD 88446    & 0.65      & 0.03  & 4     & 0.56  & 0.06  & 2     & 0.66  & 0.03    & 7     & 0.55  & 0.06  & 2     & 0.34  & 0.09  & 2      \\ 
HD 90508    & -0.08     & 0.09  & 3     & 0.11  & 0.07  & 2     & 0.06  & 0.04    & 8     & 0.18  & 0.03  & 3     & 0.17  & 0.01  & 2      \\ 
HD 109498   & 0.08      & 0.03  & 4     & 0.17  & 0.04  & 2     & 0.08  & 0.06    & 4     & 0.15  & 0.00  & 2     & 0.16  & 0.05  & 2      \\ 
HD 111367   & 0.09      & ...   & 1     & 0.14  & 0.00  & 2     & -0.04 & 0.07    & 4     & 0.06  & 0.00  & 2     & 0.02  & 0.06  & 2      \\ 
HD 135694   & 0.01      & 0.08  & 3     & 0.26  & ...   & 1     & 0.00  & 0.05    & 4     & 0.21  & 0.08  & 3     & 0.23  & 0.11  & 2      \\ 
HD 138750   & 0.03      & 0.08  & 3     & 0.12  & 0.07  & 2     & 0.04  & 0.04    & 5     & 0.04  & 0.05  & 3     & 0.01  & 0.01  & 2      \\ 
HD 140209   & -0.04     & 0.03  & 5     & 0.08  & ...   & 1     & 0.00  & 0.03    & 5     & 0.05  & 0.04  & 3     & 0.09  & 0.04  & 2      \\ 
HD 149105   & -0.09     & 0.02  & 2     & -0.1  & ...   & 1     & 0.02  & 0.03    & 4     & 0.10  & 0.11  & 2     & 0.08  & 0.05  & 2      \\ 
HD 149890   & -0.03     & ...   & 1     & -0.02 & ...   & 1     & -0.02 & 0.02    & 4     & 0.07  & ...   & 1     & 0.02  & 0.00  & 2      \\ 
HD 156617   & -0.07     & 0.02  & 4     & -0.13 & 0.04  & 2     & -0.06 & 0.04    & 5     & -0.01 & 0.02  & 3     & -0.02 & 0.05  & 2      \\ 
HD 156893   & 0.04      & 0.09  & 3     & 0.26  & 0.08  & 2     & 0.20  & 0.02    & 8     & 0.34  & 0.01  & 2     & 0.38  & 0.02  & 2      \\ 
HD 157214   & -0.08     & 0.03  & 3     & 0.13  & 0.01  & 2     & 0.04  & 0.06    & 5     & 0.27  & 0.05  & 3     & 0.28  & 0.02  & 2      \\ 
BD +40 3374 & 0.19      & 0.05  & 3     & ...   & ...   & ...   & 0.30  & 0.06    & 3     & 0.29  & 0.05  & 3     & 0.51  & 0.01  & 2      \\ 
HD 171009   & 0.05      & 0.09  & 3     & 0.27  & ...   & 1     & 0.12  & 0.06    & 6     & 0.25  & 0.06  & 2     & 0.29  & 0.03  & 2      \\ 
HD 171242   & 0.20      & 0.07  & 2     & ...   & ...   & ...   & 0.14  & 0.04    & 5     & 0.36  & ...   & 1     & 0.24  & ...   & 1      \\ 
HD 178478   & -0.04     & 0.05  & 2     & ...   & ...   & ...   & -0.04 & 0.04    & 4     & ...   & ...   & ...   & 0.19  & 0.07  & 2     \\  
HD 188326   & 0.09      & 0.05  & 3     & 0.19  & 0.01  & 2     & 0.13  & 0.04    & 7     & 0.19  & 0.05  & 3     & 0.21  & 0.04  & 2      \\ 
HD 206373   & -0.02     & 0.05  & 3     & 0.2   & 0.00  & 2     & -0.03 & 0.05    & 3     & -0.02 & 0.03  & 2     & -0.01 & ...   & 1      \\ 
HD 210483   & -0.10     & 0.09  & 4     & 0     & ...   & 1     & -0.03 & 0.04    & 6     & -0.04 & ...   & 1     & -0.02 & ...   & 1      \\ 
HD 211476   & -0.10     & 0.06  & 4     & 0.11  & ...   & 1     & 0.02  & 0.05    & 5     & 0.09  & 0.07  & 3     & 0.05  & 0.02  & 2      \\ 
HD 217511   & 0.14      & ...   & 1     & ...   & ...   & ...   & 0.20  & ...     & 1     & 0.16  & ...   & 1     & 0.14  & 0.01  & 2      \\ 
HD 219175   & -0.04     & 0.06  & 2     & 0.14  & ...   & 1     & -0.03 & 0.06    & 4     & 0.04  & 0.00  & 2     & 0.05  & 0.01  & 2      \\ 
\noalign{\smallskip}                                                                                                                        
Thin-disc stars \\                                                                                                                          
\noalign{\smallskip}                                                                                                                         
HD 115383     & -0.19   & 0.01  & 2     & 0.03  & ...   & 1     & -0.13 & 0.03    & 4     & -0.15 & 0.03  & 3     & -0.09 & 0.01  & 2      \\ 
HD 127334     & -0.20   & 0.03  & 4     & -0.14 & ...   & 1     & -0.17 & 0.05    & 6     & -0.21 & 0.04  & 2     & -0.19 & 0.06  & 2      \\ 
HD 136064     & ...     & ...   & ...   & 0.10  & ...   & 1     & 0.05  & 0.07    & 4     & ...   & ...   & ...   & 0.09  & 0.06  & 2      \\ 
HD 163989     & 0.08    & 0.03  & 3     & 0.03  & ...   & 1     & 0.04  & 0.03    & 3     & 0.06  & 0.05  & 2     & -0.01 & 0.04  & 2      \\ 
HD 187013     & -0.04   & ...   & 1     & -0.05 & 0.01  & 2     & -0.17 & 0.03    & 2     & -0.08 & 0.08  & 2     & -0.11 & 0.03  & 2      \\ 
HD 187691     & -0.16   & 0.05  & 3     & -0.10 & ...   & 2     & -0.22 & 0.04    & 4     & -0.05 & ...   & 1     & -0.07 & 0.00  & 2      \\ 
HD 200790     & -0.09   & 0.08  & 4     & -0.19 & 0.03  & 2     & -0.16 & 0.05    & 5     & -0.09 & 0.02  & 2     & -0.15 & 0.00  & 2      \\ 
HD 220117     & ...     & ...   & ...   & ...   & ...   & ...   & ...   & ...     & ...   & ...   & ...   & ...   & ...   & ...   & ...    \\ 
\noalign{\smallskip}                                                                                                                        
Thick-disc stars \\                                                                                                                         
\noalign{\smallskip}                                                                                                                        
HD 150433     & -0.08   & ...   & 1     & 0.37  & ...   & 1     & 0.12  & 0.03    & 4     & 0.34  & ...   & 1     & 0.25  & 0.03  & 2      \\ 
HD 181047     & -0.02   & 0.05  & 3     & 0.07  & ...   & 1     & -0.02 & 0.04    & 6     & 0.07  & 0.06  & 2     & 0.01  & 0.08  & 2      \\ 
HD 186411     & -0.11   & 0.05  & 4     & -0.06 & 0.13  & 2     & -0.05 & 0.04    & 6     & -0.06 & 0.02  & 2     & -0.03 & 0.01  & 2      \\ 
HD 195019     & -0.10   & 0.02  & 4     & -0.01 & 0.06  & 2     & -0.06 & 0.04    & 8     & -0.01 & 0.08  & 3     & -0.03 & 0.08  & 2      \\ 
HD 198300     & -0.03   & 0.07  & 4     & 0.28  & 0.06  & 2     & 0.07  & 0.02    & 3     & 0.25  & 0.01  & 2     & 0.12  & 0.00  & 2      \\ 

\hline
\end{longtable}
}
\end{document}